\documentclass{article}
\usepackage[left=3.5cm,right=3.5cm,top=4cm,bottom=4cm]{geometry}
\usepackage[utf8]{inputenc}
\usepackage[T1]{fontenc}
\usepackage{graphicx}
\usepackage[labelfont=bf]{caption}
\usepackage[fleqn]{amsmath}
\usepackage{textcomp}

\newcommand{\A}{\mathcal{A}}
\newcommand{\B}{\mathcal{B}}
\newcommand{\D}{\mathcal{D}}
\newcommand{\F}{\mathcal{F}}
\newcommand{\Hh}{\mathcal{H}}
\newcommand{\Ll}{\mathcal{L}}
\newcommand{\N}{\mathcal{N}}
\newcommand{\PS}{\mathcal{P}}
\newcommand{\Ss}{\mathcal{S}}
\newcommand{\T}{\mathcal{T}}
\newcommand{\X}{\mathcal{X}}

\newcommand{\DD}{\mathbb{D}}
\newcommand{\LL}{\mathbb{L}}
\newcommand{\PP}{\mathbb{P}}
\newcommand{\RR}{\mathbb{R}}
\newcommand{\TT}{\mathbb{T}}
\newcommand{\ZZ}{\mathbb{Z}}

\usepackage{amsthm}
\usepackage{amssymb,bm}
\theoremstyle{definition}

\newtheorem{example}{Example}[section]

\usepackage[pagebackref]{hyperref}
\hypersetup{colorlinks=true, citecolor=blue, linkcolor=blue, urlcolor=blue}

\renewcommand*{\backrefalt}[4]{%
    \ifcase #1 %
        [No citations.]%
    \or
        [Cited on page #2.]%
    \else
        [Cited on pages #2.]%
    \fi
}

\usepackage{enumitem}
\setlist[enumerate]{itemsep=3pt,parsep=0pt,topsep=4pt}

\title{Towards a mathematical framework for modelling cell fate dynamics}
\usepackage{authblk}  % Required for advanced author-affiliation handling
\usepackage{hyperref} % For email links
\usepackage{url}      % For formatting URLs

\makeatletter
\renewcommand\AB@affilsepx{, \protect\Affilfont}
\makeatother

\author[1,2,3]{Sean T. Vittadello}
\author[1,2]{L\'eo Diaz}
\author[3]{Yujing Liu}
\author[1,2]{Adriana Zanca}
\author[1,2]{Michael P.H. Stumpf \thanks{Corresponding author: mstumpf@unimelb.edu.au}}

\affil[1]{School of BioSciences, University of Melbourne, Australia}
\affil[2]{School of BioSciences, University of Melbourne, Australia}
\affil[3]{ARC Centre of Excellence for the Mathematical Analysis of Cellular Systems}

\date{\today}

\begin{document}

\maketitle

\begin{abstract}
    \noindent An adult human body is made up of some 30 to 40 trillion cells, all of which stem from a single fertilized egg cell. The process by which the right cells appear to arrive in their right numbers at the right time at the right place -- development -- is only understood in the roughest of outlines. This process does not happen in isolation: the egg, the embryo, the developing foetus, and the adult organism all interact intricately with their changing environments. Conceptual and, increasingly, mathematical approaches to modelling development have centred around Waddington's concept of an epigenetic landscape. This perspective enables us to talk about the molecular and cellular factors that contribute to cells reaching their terminally differentiated state: their fate. The landscape metaphor is however only a simplification of the complex process of development; it for instance does not consider environmental influences, a context which we argue needs to be explicitly taken into account and from the outset. When delving into the literature, it also quickly becomes clear that there is a lack of consistency and agreement on even fundamental concepts; for example, the precise meaning of what we refer to when talking about a `cell type' or `cell state.' Here we engage with previous theoretical and mathematical approaches to modelling cell fate -- focused on trees, networks, and landscape descriptions -- and argue that they require a level of simplification that can be problematic. We introduce random dynamical systems as one natural alternative. These provide a flexible conceptual and mathematical framework that is free of extraneous assumptions. We develop some of the basic concepts and discuss them in relation to now `classical' depictions of cell fate dynamics, in particular Waddington's landscape.
\end{abstract}

\begin{quote}
    \emph{Life can only be understood backwards; but it must be lived forwards} -- Kierkegaard
\end{quote}

\section{Introduction}
The term `cell' was originally coined by Robert Hooke, motivated by the honeycomb-like structures within cork he observed under his early microscope that reminded him of monks' cells in a monastery. The use of this term has persisted. Hooke had the luxury of seeing something new, though not so much as to invite concern over nuance. We are now in a different situation in which we can observe, probe, and measure characteristic properties of cells at astonishing resolution beyond their mere appearance. This increase in the level of attainable cellular detail means that the rough edges of our current understanding are clearly evident.

Cells are generally regarded as the smallest unit able to support life, and are firmly established as the focus and unifying theme of modern biological research. Many deep questions however remain unanswered, and furthering our understanding of cell behaviour at all levels of biological organization would provide an important foundation for both natural and synthetic biology. Fundamentally, there is no suitable framework for establishing clear definitions regarding the most basic aspects of cells and their behaviour, in particular cell type, cell identity, cell fate, and even cell state. There is, however, a diversity of perspectives in the literature, a snapshot of which is assembled in~\cite{Clevers2017What}. Confusingly, the terms cell type, cell identity, and cell state are used with seemingly all possible gradations of interchangeability, typically without clear motivation, even though they are generally considered as semantically distinct concepts. The only constant amid this uncertainty seems to be that, perhaps unsurprisingly given the subject matter, arriving at precise definitions is difficult~\cite{Dance2024What}.

Waddington's \emph{epigenetic landscape} is one of the most captivating perspectives on cell fate dynamics, and has profoundly shaped how we continue to look at cell fate and cell type more than seventy years later. The landscape is widely regarded as a conceptual framework for cell fate. Despite being elegant and evocative, however, all attempts to reconcile this visual metaphor with data and theory have revealed some shortcomings. For example, one of the more intuitive interpretations of the original metaphor assumes that development or cell differentiation can be represented by a potential energy landscape, which imposes severe limitations on the range of dynamics: cell cycle and circadian effects are difficult to accommodate; and, even randomness, which prevails at the molecular level of transcription and translation, can pose challenges. Such limitations are due in part to an overly literal interpretation of the landscape, implying a simplified perspective where a cell's phenotype can be explained in isolation from other cells and the environment. The landscape also paints a picture that is static rather than dynamic, irrespective of Waddington's intention, whereby the constantly changing and adaptive processes of cellular biology are represented as a static `geological' landscape with hills and valleys.

The study of cell fate dynamics has been motivated and limited by our historical interpretations of the subject, including the experimental techniques employed to describe and classify cells. Experimentally, there are many different aspects of cell biology that are relevant to understanding cell fate. Recent studies however often focus on molecular signatures, despite the importance of cell morphology, and many of these studies have relied solely on transcriptomic data because of technical advances rather than biological relevance. In addition to mRNA, it is known that cellular behaviour is influenced by proteins and their post-translational modifications, metabolites, lipids, and epigenetic modifications; these should be included in any description of cell state, cell type, or cell fate that aims for completeness.

Mathematically, the epigenetic landscape is often assumed to be a connected topological surface, that is, a connected 2-manifold, embedded in the higher-dimensional cell state space. While 2-manifolds may afford mathematical and conceptual simplicity, and intuitively correspond to the picture of the landscape, such a representation is unable to account for many fundamental properties of the cell. For example, cell fate determination involves complex regulatory networks which can produce very high-dimensional and complicated dynamics~\cite{Huang2005Cell}. Further, cell fate dynamics are not generally deterministic due to the influence of random processes, hence the landscape cannot be portrayed generally by a fixed and predetermined manifold. The visual metaphor of an epigenetic landscape is inherently descriptive rather than explanatory, and is not an appropriate foundation for a rigorous understanding of cell fate dynamics, a fact explicitly noted by Waddington himself~\cite{Waddington1957Strategy}.

What is needed is a mathematical framework that can represent the currently known general principles of cell fate dynamics, account for further details as our understanding of these systems improves, and guide the development of mathematical models. In this article, we therefore consider a general framework for modelling cell fate dynamics based on the theory of random dynamical systems, which integrates both change and stability into a highly adaptable conceptual and operational framework. Importantly, this framework covers all current forms of mathematical models, with both space and time either continuous or discrete, including: ordinary, partial, random, and stochastic differential equations; and, random and nonrandom difference equations.

Any attempt to model the dynamics of cell fate within the framework of a dynamical system must represent the characteristics of cell fate determination, whereby cells specialize into particular cell types with associated specific functions. Cell fate refers to the future of the cell as a cell type, and it has often been suggested that cell types correspond to attractors in dynamical systems~\cite{Huang2005Cell,Casey2020Theory}, a perspective that corresponds with the valleys in Waddington's landscape. Informally, an attractor is a subset of the state space which is invariant under the evolution of the system, and towards which certain initial states evolve asymptotically in time. Hence, cell differentiation toward a cell type can be represented in a dynamical system by the time-asymptotic evolution from an initial state toward an attractor, which represents the set of physical properties corresponding to the cell type. However, by only considering the asymptotic states of cells we may neglect the possibility that long transient dynamics, which can be so persistent that they resemble attractors, might be of fundamental importance for development~\cite{Verd2014,MacLean2018Exploring,Brackston2018Transition}.

Here we map out mathematical avenues that can help us arrive at an understanding of cell fate dynamics which is not limited by the metaphors we use to think about this cellular process. We opt for an approach that is informed by the available biological data, but that is also aware of the limitations in the data. There may be life in the landscape picture yet, and we will find it useful, directly or indirectly, to use it to shed some light on the concepts of cell state, cell type, and cell fate.

In Section~\ref{sec:background} we outline the current perspectives on cell fate, beginning with a discussion of the biology, followed by the mathematical modelling approaches, namely trees, networks, and landscapes. We also reflect on concepts such as cell type and cell state, as well as Waddington's epigenetic landscape, based on the different perspectives. Motivated by this background, we then introduce the mathematical framework for developing models of cell fate dynamics. The primary purpose of Section~\ref{sec:Represent} is pedagogical, as we would like to encourage readers unfamiliar with dynamical systems theory to engage with the main mathematical concepts of our proposed framework. In particular, we provide a relatively informal introduction to the basic concepts of dynamical systems, starting with the notions of `state space' and `evolution function' before considering the main types of dynamical system, namely autonomous, nonautonomous, and random. Section~\ref{sec:Framework} formally introduces the general mathematical framework for developing models of cell fate dynamics based on the notion of a random dynamical system. This first requires a rigorous presentation of both autonomous and nonautonomous dynamical systems. The notion of an attractor is essential for understanding the asymptotic dynamics of a dynamical system. Due to the nonautonomous nature of random dynamical systems, the corresponding attractors are time dependent, and we consider the two main types which are forward attractors and pullback attractors. We also briefly discuss the concept of long transient dynamics, which may be fundamental for understanding cell behaviour. Finally, in Section~\ref{sec:Conclusion}, we summarize our perspective and discuss some of the many ways to build on our proposed framework.

\section{Background on cell fate dynamics}\label{sec:background}
Progression towards rigorous definitions of cell fate, cell state, and cell type requires engagement with the relevant biological and, much less extensive, mathematical literature. Of relevance here is also the application of statistical mechanics. Accordingly, in this section we concisely summarize the pertinent literature, which provides motivation for the more formal discussion in subsequent sections.

\subsection{Biology}\label{subsec:bio-background}
The concept of cell fate arises in the context of the empirical observation of distinct cellular forms within multicellular organisms, and in light of the stereotypical, or repetitive, nature of development. While all cells in a given organism share the same genetic material, qualitatively different cells emerge over the course of development, in a process that repeats more or less identically across individuals. A neuron is physically and functionally distinct from a kidney cell or an epithelial cell, yet they all share the same genes. This tension between shared information and differences in phenotype (morphology, developmental processes, etc.) motivates the effort to classify these different instances, or \emph{types}, of cells. Sui Huang and Stuart Kauffman, for instance, refer to cell types as ``the most elementary `biological observable'\thinspace''~\cite{Huang2012Complex}. From this empirical perspective, cell types and their accurate classification promise to let us efficiently navigate the complexity of the space of cellular forms~\cite{Clevers2017What,Zeng2022What}.

The related but distinct concept of cell fate specifically appears to emerge from the stability of a cell's type upon terminal differentiation, which is usually coordinated via a definitive exit from the cell cycle. This stable identity of the post-mitotic cell may then be called its \emph{fate}. Development can be interpreted as the process of \emph{cell fate specification}, and we seek to identify the factors responsible for driving cells to differentiate into different types. Understanding cell fate then becomes a question about the predictability of the type that a given cell will become upon terminal differentiation, given information about its current or past states~\cite{Stumpf2017Stem}. At its most simple this process may be thought of as deterministic, meaning it is fully predictable in principle given an initial state; or more realistically as a series of reductions in uncertainty about a cell's fate as it differentiates.

While the motivation for cell types is intuitive from the morphological perspective, made possible by advances in microscopy~\cite{Gest2004Discovery}, we can now record cellular characteristics at an unprecedented level of resolution. This shift in focus at least partially motivated the use of another related notion, that of \emph{cell state}, which promises to be more precise than cell types and therefore allows us to account for the variation found within the same cell type~\cite{Clevers2017What,Casey2020Theory}. While cell state is often not clearly defined, it seems that the concept of cell fate in light of that of cell state then refers to both a cell's terminally differentiated type \emph{and} to the sequence of state transitions -- a description of how cells change states over time -- taken to reach that type.

The notion of cell state is most often defined at the molecular level, in part due to the current focus on single-cell RNA sequencing technology. This lets us focus on the (real or putative) molecular drivers of cell fate that induce so-called \emph{cell fate decisions}. These promise to identify molecular factors that drive lineage `commitment,' where a `decision' is then the `choice' of one lineage over another. This focus therefore yields definitions of cell types that are \emph{transcriptomic} in nature, that is, that are specifically defined at the restricted level of gene transcript abundance~\cite{Amini2023Evolving,Zeng2022What}.

This may not be enough, however, as we know there is a multiplicity of aspects that contribute to a cell's phenotype and to cell fate. Some of the known aspects influencing cell fate beyond gene expression are: the physical arrangement of a cell's chromatin~\cite{Buenrostro2015Singlecell}; the physical properties of the cell's microenvironment~\cite{Engler2006Matrix}; and the progression through the cell cycle~\cite{Ruijtenberg2016Coordinating}. This is far from an exhaustive list of those aspects that are known, and it is likely that other relevant factors are still unknown. Cellular function is similarly imprecisely defined and arguably determined by more than just the mRNA, or mRNA and protein census.

Crucially, the cell's internal state and its interactions with the environment are both context- and time-dependent and change over the course of differentiation. Consequently, there is currently no consensus on a rigorous definition of cell type, nor about the precise material differences between cell type and cell state~\cite{Fishell2013Neuron,Quake2021Cell,Regev2017Human,Schumacher2021Defining,Zeng2022What}. In practice, different combinations of factors will be relevant to different experimental considerations, in turn driving implicit differences in definitions.

A purely empirical perspective is limited by the available data. Even connecting different levels, from the molecular to the morphological, leaves out much of the relevant biology and the incomplete nature of such perspectives needs to be acknowledged. A natural approach for integrating different definitions of a concept is to ground them in a shared latent space, or to use mathematics to account for the (known) missing information. Connecting all of the different molecular, phenotypic, and environmental factors that can influence cellular identity can bring clarity to the meaning of cell type~\cite{Gut2018Multiplexed}. Integrating multiple data is however a known challenge~\cite{Miao2021Multiomics}, and likely requires theoretical advances in our conceptualization of cell fate. Among the existing concepts, cell state is the notion that lends itself most naturally to building this shared latent space. Such a state space may include a molecular census, but may also go further and include quantifiable properties both currently known and unknown.

This overview of the biological perspective demonstrates the need for a more general and rigorous theory of cell fate, necessarily based on experimentally validated observations incorporated within an appropriate mathematical framework. Importantly, the theory must not be limited by simplifying assumptions regarding the structure of the space of cellular identity, or by our current partial understanding of the drivers of cell fate specification. Biologically, the need for such a theory is clear. Current mathematical approaches to modelling cell fate tend to be insufficiently general and oversimplified, predominantly (and most often unnecessarily) for mathematical simplicity. We show in this article that a suitable mathematical framework exists, namely random dynamical systems, however this framework is rarely employed within the context of cell fate dynamics, despite being extensively utilized in many other fields.

\subsection{Mathematical modelling: trees, networks, and landscapes}
Here we outline the main mathematical frameworks employed for modelling cell fate in order of increasing conceptual complexity, or generality: first trees, then networks, ending with the idea of a landscape and its manifold interpretation. We denote by $\RR$, $\RR_{>0}$, and $\RR_{\ge 0}$ the set of real numbers, positive real numbers, and nonnegative real numbers, respectively.

\subsubsection{Trees}\label{subsubsec:trees}
One of the earliest uses of trees to understand cell fate is due to John Sulston and Robert Horvitz, who in 1977 reported a fully resolved \emph{cell lineage tree} of the worm \emph{Caenorhabditis elegans}, a model organism particularly well suited to this question due to its effectively deterministic development yielding invariant cell fates across individuals~\cite{Sulston1977Postembryonic}. 

Formally, a cell lineage tree is a tuple $(V,E)$ where $V$ is the set of \emph{vertices} and $E$ is the set of \emph{directed edges}, or \emph{arcs}. Cell lineage trees are therefore \emph{directed rooted trees} where all arcs point away from the \emph{root}, an arbitrarily designated vertex in $V$. This structure, also called an \emph{arborescence} or \emph{out-tree}, is a directed graph where the underlying undirected graph is connected and acyclic, and a given vertex is designated as the root.
The root of a lineage tree typically represents undifferentiated cells that may assume any fate, such as pluripotent stem cells. Terminal vertices, or \emph{leaves}, in turn denote observed terminally differentiated cell types, therefore representing the different possible fates.

The topology of a cell lineage tree imposes a time ordering that typically represents the decrease in differentiation potential from the pluripotent state to the terminally differentiated one. Specifically, if an edge exists between two vertices, only the vertex closer to the root may give rise to the one that is further away, representing the usually unidirectional process of differentiation. In other words, for an arbitrary choice of vertices in a lineage tree, those cannot be reached in the development of the organism before vertices closer to the root. In this framework, a cell `commits' to a given fate when it may only reach a single leaf, representing the cell's fate~\cite{Huang2012Complex}.

Lineage trees may represent time-dependent processes at multiple levels of resolution. In the context of lineage trees defined at the level of the organism, therefore modelling the overall process of development, the root vertex is naturally taken as the single fertilized egg cell. Lineage trees have however also been used to represent other, more specific, biological processes, such as haematopoiesis~\cite{Rommelfanger2021Singlecell}. This flexibility emphasizes that lineage trees need not describe the process they represent in full, and in fact do not in most cases; not every cell division event needs to be accounted for in a model of embryogenesis, for instance~\cite{Stadler2021Phylodynamics}.

This would otherwise imply that the tree is a \emph{binary tree}, where the \emph{out-degree}, the number of arcs directed away from a vertex, of all non-terminal vertices is exactly two. In the context of cell fate, a binary tree corresponds to the partitioning of the set of leaves that may be reached from any given vertex into exactly two sets. This partitioning is what allows us to refer to cell fate `decisions,' or the `choice' of one set of fates over another associated to any outgoing arc.

There are cases where binary trees are not the most relevant structure. For lineage trees representing haematopoiesis for instance, common myeloid progenitor cells may assume more than two fates, giving rise to either megakaryocytes, erythrocytes, mast cells, myeloblasts, etc., and not every vertex will have two outgoing arcs. Additional cell types may however be defined such that the lineage tree becomes binary, even without accounting for all cell divisions~\cite{Rommelfanger2021Singlecell}. Lineage trees that represent all cell division events can also be formulated. Such lineage trees provide more granular information on the path to terminally differentiated states, which would necessarily make such a tree a binary tree. There are also cases where the very concept of cell lineage trees becomes less relevant, such as when dedifferentiation occurs, which can introduce reticulations.

What we have described so far are \emph{observational} structures that recapitulate our existing knowledge of differentiation processes. To create a \emph{predictive} lineage tree, one may turn to processes on tree-like structures, such as branching processes or agent-based modelling, which may be probabilistic. Incorporating mechanisms into process models may go some way to explaining why certain cell fates arise, that is, help to identify molecular events that drive lineage commitment, for instance. This kind of reasoning however goes beyond the structures described by trees and requires exploring both intracellular and extracellular interactions.

\subsubsection{Networks}\label{subsubsec:networks}
While trees naturally lend themselves to talking about cellular lineages and differentiation processes, a more complex description is needed to account for the multiplicity of molecular events at the subcellular level that may affect a cell's phenotype. A first step towards this more detailed view is realized in the framework of the \emph{gene regulatory network}, or GRN. This gene-centric perspective builds on the dynamics of gene expression as regulated by \emph{transcription factors}, proteins that modulate the rate of gene transcription by binding specific nuclear regions called \emph{regulatory sequences}~\cite{deJong2002Modeling}. Genes may then be thought of as the vertices of a graph whose edges therefore represent `regulatory' gene-gene interactions, such as transcription factors binding events.

Since the GRN framework is defined at the genomic level, the \emph{state} of a given cell is defined in terms of its genes, usually as the copy number or concentration of expressed gene products. Assuming $N$ genes, we can represent the cell state $s$ as an $N$-dimensional vector $s := (x_1,\ldots,x_N)$, where each element is the molecular abundance of the corresponding gene. The set of all possible states $s$ is the \emph{state space} $S \subseteq \RR_{\ge 0}^N$. Each state $s \in S$ represents an instantaneous snapshot of the dynamic behaviour of a cell's gene expression processes, referred to as a gene expression \emph{pattern} or \emph{profile} to indicate it is a composition of individual gene product abundances.

The amenable mathematical properties of this network-based formalism allow us to study gene expression dynamics in a tractable way. The analysis is often carried out using differential equations, and, at its most general, the global dynamics arising from a GRN may be written as
\begin{equation}\label{eq:network-dynamics}
    \frac{\mathrm{d}s(t)}{\mathrm{d}t} = F(s(t)),
\end{equation}
where $s(t)$ is the time-dependent vector describing the state $s$ of a cell at some time $t$, and $F$ is a function describing both the structure (e.g. as an interaction matrix) and mathematical formulation of the gene-gene interactions encoded by the GRN~\cite{Huang2012Complex,Casey2020Theory}. This formulation thus treats any GRN as a dynamical system. Since GRNs are meant to represent the underlying biologically-occurring molecular reactions, edges in the network are meant to represent true biological interactions: in practice, those are only the ones that have been observed empirically, as well as those that are possible. The precise form taken by the function $F$ depends on the level of detail or biological faithfulness required, and different approaches to defining it correspond to specific network-based modelling formalisms.

In the network framework, cell types emerge as \emph{stable modes} of gene expression, a term which informally refers to some degree of stationarity in the dynamics~\cite{Kauffman1969Metabolic}. This rests on a phenomenological identification between the observed persistence of cellular characteristics at the phenotype-level and stability at the level of molecular reaction dynamics. This mapping is intuitive and facilitates mathematical analysis as it is defined at the GRN level, which thus in theory becomes sufficient to understanding the process of cell fate determination.

Taking this line of argument one step further then yields an identification of the gene expression pattern at steady-state with an \emph{attractor} in a high-dimensional state space~\cite{Huang2005Cell}. The intuitive idea that distinct cell types should correspond to different attractors directly follows from this identification. That steady-states specifically correspond to cell types is however only a first approximation: knowledge of the existence of a fixed point is on its own not enough to explain the stability of a phenotype, which would require the fixed point to have self-stabilising properties~\cite{Huang2012Complex}. While this holds for attractors by definition (attractors are self-stabilising with respect to `small' perturbations), any fixed point is not necessarily an attractor under the dynamical systems theory framework; a specific example would be an unstable fixed point, which is instead referred to as a \emph{repellor}. It is also important to note that there exist many kinds of attractors, with some representing subtle and complex dynamical behaviour that would be of relevance in understanding cell fate with more nuance~\cite{Casey2020Theory}. Nonetheless, this perspective allows one to explicitly frame the process of cell fate determination as the dynamics of cells moving in and out of attractors in gene expression space, an effective reduction that makes the problem both mentally and mathematically tractable.

One possible representation of GRNs is as Boolean networks~\cite{Kauffman1969Metabolic}. Here a Boolean network is defined as being composed of $N$ discrete elements (genes), each receiving $K$ Boolean inputs, representing gene expression regulation. This representation thus describes `binary' genes, either `on' (expressed) or `off' (repressed) at a given time. Boolean maps are constructed to integrate $K$ inputs into a binary response describing gene states (the number of possible maps to choose from increases exponentially, i.e. $2^{2^2}=16$ maps for $K=2$, $2^{2^3} = 256$ for $K=3$, etc.). Dynamics are then obtained under the coordinated action of the Boolean maps over discrete time steps. Specifically, at each step, each Boolean map updates the local state of the genes it connects, thus making the dynamics Markovian. This procedure effectively defines an \emph{evolution function} under the dynamical systems theory framework, a notion we will develop in more detail in Section~\ref{sec:Represent}.

This approach allows us to analyse the dynamical properties of such genetic Boolean networks systematically following their construction, which incorporates an element of randomness. Practically, random Boolean networks are constructed by choosing $K$ inputs in $N$ and randomly assigning Boolean maps with $K$ inputs to each of the $N$ binary genes, the resulting random network then behaving deterministically~\cite{Kauffman1969Metabolic}. Note that a random Boolean network generates a discrete time and discrete state random dynamical system, as developed in Section~\ref{sec:Represent}.

A natural notion here is that of a \emph{cycle}, a sequence of binary gene states that, once entered, is never left under deterministic dynamics. Of interest here is the study of the dynamics of the network under perturbations of its states; this lets us quantify the likelihood of the system returning to a cycle following perturbations, or of transitioning to another cycle if the network admits more than one. Cycles may here be identified with cell types, and a natural extension is then to investigate the number of cycles a given Boolean network is capable of supporting. Note that a cycle constitutes an \emph{invariant set}, a notion that partially overlaps with that of an attractor in dynamical systems theory. This makes the use of cycles in random Boolean networks analogous to that of attractors in the GRN framework, as demonstrated by the parallel identification of cycles (from the Boolean network perspective) and of fixed points (from the GRN perspective) with cell types, although this is not made explicit in~\cite{Kauffman1969Metabolic}.

Chemical reaction networks are another network-based formalism that have the advantage to let the dynamics, $F$ in Equation~\eqref{eq:network-dynamics}, be formulated with more biological details~\cite{Aris1965Prolegomena}. Molecular \emph{species} (usually representing gene products) can form \emph{complexes} that interact and transform along \emph{reaction channels}. We write $r: c \to c'$ to indicate that complex $c$ turns into complex $c'$ along reaction $r$. Collecting reactions as the edges of a digraph yields a representation of GRNs as two graphs, one a bipartite graph indicating the way species compose into complexes, the other a directed graph representing how complexes interface along reaction channels~\cite{Gatermann2001Counting}.

The chemical reaction network formalism yields a dynamical update rule, the \emph{species formation rate function} $f$, defined generally at state $s(t)$ as
\begin{equation}\label{eq:species-formation-rate-function}
    f(s(t)) := \sum_{c \to c' \in R} \lambda_{c \to c'}(s(t)) (c' - c),
\end{equation}
where $R$ is the set of all reactions considered, $\lambda_{c \to c'}$ is a reaction-specific \emph{rate function}, and $c' - c$ denotes the net change in species state composition under the action of each reaction. The specific form of the rate function may be chosen according to the level of detail required. In particular, under the commonly made assumption of \emph{mass-action}, which states that the rate at which a reaction occurs is proportional to the probability of collision of species in complex $c$, the form of $\lambda_{c \to c'}$ is fixed for all reactions and Equation~\eqref{eq:species-formation-rate-function} becomes
\begin{equation}
    f(s(t)) = \sum_{j = 1}^{|R|} k_j \prod_{i = 1}^{N} x_i^{\nu_{ij}} (\nu_{ij}' - \nu_{ij}),
\end{equation}
with $k_j$ a reaction-specific chemical parameter in the set of positive real numbers $\mathbb{R}_{>0}$, and $\nu_{ij}$ (respectively $\nu'_{ij}$) denoting the number of times each species $x_i$ appears in complex $c$ (respectively $c'$) in reaction $j$, subtracted element-wise.

More detailed forms of the network dynamics that explicitly consider the randomness inherent to chemical kinetics may also be formulated~\cite{Gillespie1976General}. While a sole focus on genes as a structural unit is limited~\cite{FoxKeller2005Century}, GRNs are from this perspective a flexible and powerful framework allowing for the study of gene expression dynamics at different levels of abstraction, depending on the precise interpretation of the rate functions. More generally, the network formalism forms the basis of the landscape idea, which allows us to go one step further and explicitly connect molecular processes to phenotype.

\subsubsection{Landscapes}
Waddington's epigenetic landscape aims to provide an intuitive picture of the dynamics of cell differentiation processes~\cite{Waddington1957Strategy}. Under this evocative metaphor, the progressive differentiation of pluripotent stem cells into terminally differentiated cells is likened to balls rolling down a landscape of valleys separated by ridges. Here, valleys symbolize the paths followed by cells under stereotypical development, each ending in distinct cell types -- the cells' fate -- while ridges represent barriers separating the different fates.

This picture naturally lends itself to a physical interpretation, where stem cells represent a high-energy undifferentiated state that progressively loses potential as the metaphorical cell rolls towards the valleys of the landscape. Waddington's concept of \emph{canalization} then describes the tendency of developmental processes to return to the path expected under stereotypical development following perturbations. This stability visually translates to the topology of the landscape, where `canalized' paths are drawn with steep hillslopes that represent this compensatory ability.

This conceptual model integrates trees, and more specifically lineage trees in developmental biology, as discussed above in Section~\ref{subsubsec:trees}; here, each branching point represents a potential `choice' leading to different cell fates. The valleys of the landscape dictate the structure of the tree, where branching points then represent potential `decisions' or bifurcations~\cite{Stadler2021Phylodynamics}.

Waddington envisioned his landscape to be grounded in epigenetic space: cell fate determination may be represented by changes in gene expression. In his metaphor, genes are represented by pegs attached to the ground, and the influence of each gene on the structure of the landscape by strings leading out of these pegs, pulling and folding the surface of the landscape into valleys and ridges. The set of pegs and strings forms a network (a GRN for instance, as discussed above in Section~\ref{subsubsec:networks}) that regulates developmental dynamics and therefore should define the structure of the landscape. Note, however that the landscape being defined at the (epi)genetic level primarily focuses on gene expression, excluding other potential factors influencing cell fate, as discussed above in Section~\ref{subsec:bio-background}.

More formally, the state of a cell can be expressed as a vector $x \in \mathcal{X}$ containing the abundances of all relevant molecular species within some gene expression state space $\mathcal{X}$, which is defined a priori as the set of all possible configurations of molecular abundance. Molecular abundances are effectively discrete -- there is a finite number of copies of a given molecule -- but are usually understood in terms of concentrations. A state $x$ can therefore be understood as a continuous variable, and the space $\mathcal{X}$ can be represented by a subset of $\RR_{\ge 0}^d$, where $d$ is the dimension of $x$.
\par
We can then describe the change in $x$ with respect to time $t$ through e.g. a stochastic differential equation of the type
\begin{equation}\label{eq:SDE}
    \mathrm{d}x = f(x;\theta,t) \mathrm{d}t + g(x;\theta,t) \mathrm{d}W_t.
\end{equation}
Here, $f(x;\theta,t)$ and $g(x;\theta,t)$ represent the deterministic and stochastic components of the dynamics, respectively, also known as \emph{drift} and \emph{diffusion}, where $\theta$ denotes the parameters of the system. The term $\mathrm{d}W_t$ with $E[\mathrm{d}W_t] = 0$ and $E[\mathrm{d}W_t\mathrm{d}W_s] = \epsilon\delta(t-s)$ is the so-called Wiener process increment that instantiates the randomness, where $\epsilon$ is the noise strength. The stochastic component is considered here to account for internal fluctuations at the molecular level or small environmental perturbations, the contribution of which to cellular behaviour is important to consider at the cellular level.
\par
When the system is purely deterministic with $g(x;\theta,t) = 0$, the stable fixed points of the system denoted by $x^*$ are given by
\begin{equation}
    f(x^*;\theta,t) = 0,
\end{equation}
where the Jacobian of $f(x^*;\theta,t)$ only admits negative eigenvalues. For simplicity, consider the system constructed under a scalar potential function, $U(x;\theta,t)$, representing the landscape, with $f(x;\theta,t) = -\nabla U(x;\theta,t)$. The stable points of the system are the local minima of the potential function $U(x;\theta,t)$ and cell fates are thus determined by these local minima, or `valleys' in the landscape.
\par
Let $\mathcal{V} = \{V_1,\ldots,V_n\}$ be the set of all distinct cell fates associated with a given landscape. A \emph{cell fate transition} then denotes the change in state $x$ from one valley $V_i$ to another $V_j$,
\begin{equation}
    x\in V_i \longrightarrow x\in V_j,
\end{equation}
with $i$, $j \in \{1,\ldots,n\}$ and $i\ne j$, as determined by the dynamics of the system (Equation~\eqref{eq:SDE}). We are thus able to qualitatively describe fate transitions on the landscape in terms of quantitative differences at the dynamics level. This framework lets us explore fate transitions in biologically-relevant cases, including reprogramming, and promises to provide insight into the biological mechanisms underpinning these events.

Capturing cells during the transition process has, however, been challenging, in part due to the incoherence of dynamics outside the `canalized' paths, as driven by external fluctuations that can reconfigure the landscape. This has led to the development of the idea of \emph{transition states}~\cite{Moris2016Transition,Brackston2018Transition,MacLean2018Exploring}, a concept which allows us to account for the observed variability in experimental data while staying on the landscape.

One success of Waddington's landscape has been to shed light on the fundamental principles of differentiation, as enabled by the development of single-cell RNA sequencing technology. This advance has indeed made available quantitative transcriptomic data which, in the context of the landscape as a conceptual framework linking genotype to phenotype, means we can identify the molecular mechanisms driving cellular biology, thereby providing insight into cell fate decision-making systems. This has conferred the landscape with immediate practical applications, making the concept more relevant than ever as we now are able to reconstruct landscapes from experimental data~\cite{Liu2024Approximate}.
\par
The dynamical behaviour of the landscape can be effectively characterized in low-dimensional gene expression spaces. However, in biologically realistic cases where more than thousands of genes are interacting, it is difficult to solve the resulting high-dimensional equations and the landscape has limited practical utility for understanding the underlying system in these cases. Additionally, certain dynamical behaviours relevant to modelling biological systems cannot be accounted for under the landscape metaphor, such as limit cycles which are relevant in the context of the cell cycle. The landscape metaphor further assumes an effective independence of cells since it cannot account for environmental influences on cell fate specification. This is a major limitation in light of existing biological knowledge on the importance of cell--cell communication in the form of signalling, which is crucial in establishing a cell's terminal identity in a number of cases~\cite{Rommelfanger2021Singlecell}. How to integrate the internal cellular milieu and a cell's microenvironment into a landscape representation is however not obvious.
\par
Most modelling efforts around the concept of the epigenetic landscape seem to focus on attempting to make the original intuitions precise. Beyond the obvious limitations in formalising the simplified visual representation of the landscape, it is natural to ask why there has been so little effort going into generalising the ideas introduced by Waddington. A natural extension of the landscape would, for instance, be a manifold: this would allow us to represent the high-dimensional space of cell states, and be a more flexible and realistic representation of development than lineage trees~\cite{Wagner2020Lineage}. While the idea of using manifolds is sometimes mentioned, it does not seem to have been explored rigorously or extensively. Of importance here is that the use of the term `manifold' in the context of single-cell studies is distinct from, and not equivalent to, the mathematical concept~\cite{Miao2021Multiomics}. Recently, Rafelski \& Theriot proposed a holistic perspective on the landscape as a high-dimensional state manifold indexed by four categories of cellular observables that stabilize each other, as if connected by springs~\cite{Rafelski2024Establishing}. While this does not provide a rigorous treatment of the use of manifolds either, it is a welcome and needed conceptual improvement towards more general and nuanced perspectives.

\subsection{Statistical mechanics}
Statistical mechanics~\cite{Sethna2006Statistical} aims to explain the macroscopic behaviour of a system in terms of its microscopic dynamics, defined on sets of \emph{microstates}. Our understanding of cell fate is at a point where we have a wealth of knowledge about observable aspects of development but comparatively little integrated insight into the molecular, physiological, and environmental factors driving cell fate specification. The tension between phenotype-level descriptions and fluctuating molecular dynamics is proving to be a fertile ground for applying tools and concepts from statistical physics: microstates may be related to molecular abundances and structural arrangements we usually refer to as cell state, while macrostates can be defined more flexibly, with phenotypes -- cell types for instance -- being suitable candidates~\cite{Guillemin2020Nonequilibrium}.

This logic is already in use in the landscape metaphor, where molecular quantities are linked to phenotypes. The framework of statistical mechanics makes this formal, and it has been applied to relate population-level properties to trajectories at the microscopic level. More specifically, instead of focusing on a single cell's trajectory on a landscape, we consider a population of cells that are originally in the same state $x$ for some initial time $t$. The resulting \emph{population profile} $\rho(x,t)$ is a probability density function describing the likelihood of the possible molecular configurations of the system. Since the potential function $U(x;\theta,t)$ representing the landscape itself does not account for stochasticity, the nonlinear or nonlocal effects within the framework, which would amount to drift and diffusion terms, can only be obtained from the population profile.

In this setting, cell fates correspond to the local maxima of the probability density function $\rho(x,t)$ when the system admits a stationary state, where $\rho(x,t) \to \rho(x)$, as $t \to \infty$. Under equilibrium conditions, the stationary state $\rho(x)$ relates to the landscape by
\begin{equation}
    U_q(x) \propto -\ln\rho(x), 
\end{equation}
where $U_q(x)$ is known as the \emph{quasi-potential landscape}~\cite{Coomer2022Noise} of the system. In this framework, the local maxima in $\rho(x)$ are mapped into local minima in $U_q(x)$, and cell fates are therefore defined by these minima or `valleys' in the quasi-potential landscape. Each fate here corresponds to a neighbourhood around a valley, as identified by
\begin{equation}
    \nabla U_q(x^*) = 0, \quad \nabla\cdot\nabla U_q(x^*)<0.
\end{equation}
\par
Another possible use of statistical mechanics in this context is to study the pathways of cell fate determination. The presence of random fluctuations in the system can lead to many possible transitions from one cell fate to another~\cite{Zakine2023MinimumAction}. However, not all of these paths are equally likely. When the noise term $g(x;\theta,t) = 0$ or the noise strength $\epsilon \to 0$, the ensemble of paths concentrate around the \emph{least action path}~\cite{Brackston2018Transition}. This can be viewed as mapping all the possible microscopic trajectories from one state to another to a macroscopic ensemble, where the least action path is the most likely transition.
\par
Most of our knowledge of physical systems is established in equilibrium conditions, where microscopic behaviours are time-reversible and transitions are therefore symmetric. This assumes that, given two fates $V_i, V_j \in \mathcal{V}$, the transition from $V_i$ to $V_j$ is identical to that from $V_j$ to $V_i$. Biological systems such as cells are, however, out of equilibrium, mainly due to the permeability of the cell membrane that allows for transport of energy and matter as cells respond and adapt to their environment. The forward and backward paths can only coincide if the deterministic component is purely a gradient system, and the stochastic component is constant~\cite{Wang2011Quantifying}, i.e. if
\begin{equation}
    \text{$f(x;\theta,t) = - \nabla U(x;\theta,t)$ \, and \, $g(x;\theta,t) = \text{constant}$.}
\end{equation}
This, however, will not hold under plausible biological conditions, where transitions are instead time-irreversible. This means that the least action path from one fate $V_i$ to another $V_j$ is distinct from the path from $V_j$ to $V_i$, or, more specifically, that the path in gene expression state space from one fate to another is not equivalent to its inverse.
\par
We may, however, obtain non-equilibrium properties from knowledge of equilibrium properties. For example, when the system is in equilibrium, we know that it returns to the stationary point with a force that is proportional to the amplitude of the fluctuations. Thus the variability or magnitude of fluctuations, here $g(x;\theta,t)$, is related to the drift force, $f(x;\theta,t)$. This key result is known as the \emph{fluctuation-dissipation theorem}~\cite{Attard2012Nonequilibrium} and allows us to consider some non-equilibrium processes in terms of equilibrium dynamics.

\section{Representing cell fate dynamics as a dynamical system}\label{sec:Represent}
In this section we provide a motivating discussion of the fundamental concepts for representing cell fate dynamics as a dynamical system, based on current understanding of cell systems as outlined in Section~\ref{sec:background}. There are, broadly, two main concepts: the \emph{state space} of all relevant cellular properties; and, the \emph{evolution function} on the state space which describes the cellular dynamics. Dynamical systems theory is fundamental for developing models of systems that evolve in time, which are ubiquitous in nature, and consequently there are myriad monographs on the subject. An example that concisely covers the relevant theory for this article is~\cite{Kloeden2011Nonautonomous}. We begin with a description of the sets and algebraic structures that we henceforth employ.

\subsection{Basic notation and definitions: sets and algebraic structures}

\subsubsection{Sets}
Denote by $(\RR,+,\cdot,\le)$ the ordered field of real numbers with standard operations on $\RR$ of addition $+$ and multiplication $\cdot$, and standard nonstrict (resp. strict) total order $\le$ (resp. $<$). The subsets of nonnegative real numbers $\RR_{\ge 0}$ and positive real numbers $\RR_{> 0}$ have the inherited operations and orders.

Denote by $\ZZ$ the set of integers, and by $(\ZZ,+,\cdot,\le)$ the ordered integral domain of integers with standard operations on $\ZZ$ of addition $+$ and multiplication $\cdot$, and standard nonstrict (resp. strict) total order $\le$ (resp. $<$). The subsets of nonnegative integers $\ZZ_{\ge 0}$ and positive integers $\ZZ_{>0}$ have the inherited operations and orders.

While we use the same notation for the operations and orders on the sets $\RR$ and $\ZZ$, as well as their subsets, it will be clear from context which we intend.

For $n \in \ZZ_{> 0}$ denote $[n] := \{\, m \in \ZZ_{> 0} \mid m \le n\,\}$. For any set $A$ we denote by $2^A$ the power set of $A$. If $(P,\le)$ is a partially ordered set then let $P^{(2,\le)}$ be the relation over $P$ given by $P^{(2,\le)} := \{\, (p,q) \in P^2 \mid p \le q \,\}$.\\

\subsubsection{Algebraic structures}
\textbf{Semigroup, monoid, group:} A \emph{semigroup} $(S,+)$ consists of a nonempty set $S$ and an associative binary operation $+$ on $S$. A \emph{monoid} is a semigroup with an identity element, denoted $e$ or $0$. A \emph{group} is a monoid in which every element is invertible. We use $+$ to denote the binary operation as we will always write these structures additively. Note that references to semigroups also account for monoids and groups, unless otherwise specified. A semigroup $(S,+)$ is \emph{commutative} if the binary operation is commutative.

\textbf{Totally ordered semigroup:} A \emph{totally ordered semigroup} $(S,+,\le)$ is a semigroup $(S,+)$ with a total order $\le$ on $S$ that is compatible with the binary operation (translation-invariance property): $s \le t$ if and only if $u + s \le u + t$ if and only if $s + u \le t + u$, for all $s$, $t$, $u \in S$. The \emph{strict order} $<$ associated with the \emph{nonstrict order} $\le$ is defined by $s < t$ when $s \le t$ and $s \ne t$ for all $s$, $t \in S$.

\textbf{Cones:} Let $(S,+,\le)$ be a totally ordered semigroup. An element $s \in S$ is \emph{positive} when $s < s + s$, and the set $S^+$ of all positive elements in $S$ is the \emph{positive cone} of $S$.  The positive cone is a subsemigroup of $S$ with the inherited binary relation: to see that $S^+$ is closed under $+$, if $s$, $t \in S^+$ then $s < s+s$ implies $s+t < (s+s)+t$ implies $t < s+t$, and $t < t+t$ implies $s+t < s+(t+t)$ implies $s < s+t$, so $s+t < (s+t)+t < (s+t)+(s+t)$, and hence $s+t \in S^+$.

Similarly, $s \in S$ is \emph{negative} when $s + s < s$, and the set of all negative elements is the \emph{negative cone} $S^-$ of $S$, which is a subsemigroup of $S$. If $S$ has an identity $e$ then $s \in S$ is positive (resp. negative) if and only if $e < s$ (resp. $s < e$). Then $s \in S$ is \emph{nonnegative} (resp. \emph{nonpositive}) when $e \le s$ (resp. $s \le e$), with the corresponding cones defined accordingly. If $(S,+,\le)$ is a totally ordered group then $S = S^- \cup S^+$.

\textbf{Topological semigroup:} A \emph{topological semigroup} $(S,+,\T)$ consists of a semigroup $(S,+)$ and a topological space $(S,\T)$ such that the operation $+ \colon S \times S \to S$ is jointly continuous. If $(S,+)$ is a group then we further assume that the inversion map $s \mapsto -s$ on $S$ is continuous.

\textbf{Measurable semigroup:} A \emph{measurable (semi)group} $(S,+,\Ss)$ consists of a (semi)group $(S,+)$ and a measurable space $(S,\Ss)$, where $\Ss$ is a $\sigma$-algebra over $S$, such that the operation $+ \colon S \times S \to S$ is $(\Ss \otimes \Ss,\Ss)$-measurable (jointly measurable), and if $(S,+,\Ss)$ is a group then additionally the map $s \mapsto -s$ on $S$ is $(\Ss,\Ss)$-measurable. If $(S,+,\Ss)$ is a measurable (semi)group and $S$ has a topology then, unless otherwise specified, $\Ss$ is assumed to be the Borel $\sigma$-algebra generated by the open sets.

\textbf{Standard examples:} $(\RR,+,\le)$ and $(\ZZ,+,\le)$ are totally ordered commutative groups, $(\RR_{\ge 0},+,\le)$ and $(\ZZ_{\ge 0},+,\le)$ are totally ordered commutative (non-group) monoids, and $(\RR_{> 0},+,\le)$ and $(\ZZ_{> 0},+,\le)$ are totally ordered commutative (non-monoid) semigroups. Moreover, $\RR_{\ge 0}$ (resp. $\RR_{>0})$ is the nonnegative (resp. positive) cone of $\RR$, and similarly for the sets of integers. With respect to the standard topologies these groups/monoids/semigroups are topological.

\subsection{State variable, state, and state space}
The cell interior is partitioned from the extracellular environment by the selectively permeable cell membrane, so the cell is a dissipative system: it is thermodynamically open and exists in a state of thermodynamic nonequilibrium. Cell behaviour is influenced by the chemical, mechanical, and morphological characteristics of not only the cell itself, but also of the cell's microenvironment: the local extracellular environment with which the cell interacts and that directly or indirectly influences the behaviour of the cell; in general it consists of biochemical signalling molecules, the extracellular matrix, and nearby cells. Therefore, cell behaviour generally depends on the interrelated cell--microenvironment system, though we can maintain the delineation of the cell as a fundamental unit by regarding the cell and the microenvironment as two separate, though coupled, systems rather than as a single system.

Our aim is to describe cell behaviour mathematically in terms of relevant quantitative chemical, mechanical, and morphological properties of the cell--microenvironment system, where each such system property is a time-dependent \emph{state variable}. Assume there are $n \in \ZZ_{>0}$ real-valued state variables denoted $x_i$, or $x_i(t)$ at time point $t$, for $i \in [n]$. Note that the state variables may be either continuous or discrete. A \emph{state} $x(t)$ of the system corresponds to the values of all state variables at $t$, given by the coordinate vector $x(t) := (x_1(t), x_2(t), \ldots, x_n(t))$ in real $n$-space $\RR^n$. The \emph{state space} $X \subseteq \RR^n$ consists of all possible states of the system.

We need to ensure that the chosen state variables are sufficient to describe the evolution of the system, which may be nontrivial due to the complexity of the cell--microenvironment system. Comparing predictions from mathematical models with experimental observations may reveal such issues. To minimize $n$ we only consider the state variables that are relevant for the system of interest. This may mean that the state space $X$ corresponds to a subsystem of the cell--microenvironment system.

The specific realization of cell state may vary both between cells and over time. While homeostatic regulation generally ensures that many internal properties of the cell are relatively consistent, fluctuations in the external environment may result in changes in the relevant external properties, and in turn some internal properties. We may describe the cell state at the level of detail we require, mitigated by the ability to experimentally quantify the corresponding state variables. At the most detailed, and impractical, level, the cell state includes the state variables for all molecular species in the cell, including morphology and spatial structure. More practical descriptions of cell state may be restricted to levels of gene expression, quantified functional properties of the cell, and extracellular drivers of cell behaviour.

\subsection{Evolution function}
Let $X$ be a state space corresponding to a cell--microenvironment system. We would like to model the dynamics of cell fate as state transitions, or evolution. For a state transition in $X$ from $x$ to $y$, we say that $x$ is the \emph{initial state} and $y$ is the \emph{final state}. We first consider appropriate notions of time and duration. While biological systems always evolve forwards in time, for modelling purposes it can be useful to allow evolution backwards.

\subsubsection{Time set and duration set}
Notions of time are fundamental to dynamics, so we begin with the relevant definitions. A nonempty set $\TT$ is a \emph{time set} if  $(\TT,+,\le)$ is a totally ordered subsemigroup of $(\RR,+,\le)$, and then each element of $\TT$ is a \emph{time point}. Note that $\TT$ parameterizes, and does not have to correspond with, physical time. If $s$, $t \in \TT$ and $s < t$ then we say that $t$ is forward in time with respect to $s$, and $s$ is backward in time with respect to $t$. In particular, if $\TT \subseteq \ZZ$ then $\TT$ is \emph{discrete time}, and if $\TT \subseteq \RR$ is an interval then $\TT$ is \emph{continuous time}. We call $\DD := \TT^{(2,\le)}$ the \emph{duration set} corresponding to $\TT$, and each element $(s,t) \in \DD$ is a \emph{duration}. Note that $\DD$ is a commutative semigroup under coordinate-wise addition. Further, $\LL := \{\, t-s \mid (s,t) \in \DD \,\}$ is the \emph{duration-length set} corresponding to $\TT$, and  $(\LL,+,\le)$ is a totally ordered submonoid of $(\RR,+,\le)$.

\subsubsection{Autonomous deterministic systems (time-point formulation)}
Autonomous systems depend only on the duration length, so are independent of the actual initial time point assuming the same initial state. This means that there are no external influences that produce differences in system behaviour from a given initial state at different starting times. Biological systems generally interact with their surrounding environment so are typically nonautonomous, however under appropriate conditions a nonautonomous system can sometimes be approximated by an autonomous system. The following four properties characterize an autonomous deterministic system.

\textbf{Determinism:} A system is \emph{deterministic} if any final state is uniquely determined by the initial state alone, so we can describe the state transitions with a function $\eta \colon \DD \times X \to X$, called the \emph{evolution function}. If the initial state is $x \in X$ at the initial time point $s \in \TT$ then at the final time point $t \in \TT$ the final state is $\eta((s,t),x) \in X$.

\textbf{Autonomy:} A system is \emph{autonomous}, or \emph{time-translation invariant}, or \emph{time invariant}, if the dynamics of the system depend only on the duration length and not on the initial and final time points. Autonomy is represented with the evolution function $\eta$ by assuming it satisfies \emph{time-translation invariance}, that is, $\eta((s,t),x) =\eta((s+r,t+r),x)$ for all $((s,t),x) \in \DD \times X$ and $r \in \TT$. Therefore, with the same initial state $x \in X$ the system will always reach the same final state after the same duration length, irrespective of the initial time point.

\textbf{Identity:} If the initial and final time points are the same, so the duration length is zero, then we specify no change in state since we assume the relevant physical processes always occur over nonzero duration lengths. Therefore, $\eta((s,s),x) = x$ for all $x \in X$ and $s \in \TT$.

\textbf{Causal relation:} Suppose $((s,t),x) \in \DD \times X$ and there exists $r \in \TT$ with $s < r < t$, so that the duration $(s,t)$ is divisible into the durations $(s,r)$ and $(r,t)$. Note that divisibility of durations depends on $\TT$: for example, if $\TT = \RR$ then all nontrivial durations are divisible, however if $\TT = \ZZ$ then durations of length one are not divisible. The divisibility of durations corresponds to the \emph{causal relation}, which describes the connection between the initial and final states through the intermediate states. The causal relation is represented with the evolution function by requiring $\eta((s,t),x) = \eta((r,t),\eta((s,r),x))$ for all $((s,t),x) \in \DD \times X$ and $r \in \TT$ with $s < r < t$. That is, the transition from the initial state $x$ at initial time point $s$ to the final state $\eta((s,t),x)$ at final time point $t$ is the same as, by causation, the transition from the initial state $x$ at initial time point $s$ to the intermediate state $\eta((s,r),x)$ at time point $r$ followed by the transition from $\eta((s,r),x)$ at time point $r$ to the final state $\eta((r,t),\eta((s,r),x))$ at time point $t$.

\subsubsection{Autonomous deterministic systems (duration-length formulation)}\label{subsubsec:autonomous_duration}
Since autonomous systems are independent of time points we can describe the evolution of the system explicitly in terms of duration lengths. Define the evolution function $\theta \colon \LL \times X \to X$ by $\theta(l,x) := \eta((s,t),x)$ for some $((s,t),x) \in \DD \times X$ with $l=t-s$. Note that $\theta$ is well defined: if also $((u,v),x) \in \DD \times X$ with $l=v-u$ then, assuming without loss of generality that $s<u$, we have $\eta((s,t),x) =\eta((s+(u-s),t+(u-s)),x) =\eta((u,v),x)$, where the first equality follows from the time-translation invariance of $\eta$, and the second equality from $t-s=v-u$. Autonomy and identity follow immediately from the corresponding properties of $\eta$. For the causal relation, suppose $l$, $m \in \LL$ and $x \in X$. Let $(s,t)$, $(t,u) \in \DD$, which we can choose with the common time point $t$ by time-translation invariance, such that $l=t-s$ and $m=u-t$. Then $\theta(l+m,x) =\eta((s,u),x) =\eta((t,u),\eta((s,t),x)) = \theta(m,\theta(l,x))$, where the first equality follows since $l+m = u-s$, and the third equality holds by the causal relation of $\eta$. Note that we can obtain $\eta$ from $\theta$ by defining $\eta((s,t),x) := \theta(t-s,x)$ for $((s,t),x) \in \DD \times X$.

\textbf{Semigroup property:} Defining $\theta_l \colon X \to X$ by $\theta_l(x) := \theta(l,x)$ for all $l \in \LL$ and $x \in X$, we can express the causal relation in the form $\theta_{l+m} = \theta_m \circ \theta_l$ on $X$ for all $l$, $m \in \LL$, which is called the \emph{semigroup property} since the functions $(\theta_l)_{l \in \LL}$ form a semigroup (in fact a commutative monoid with identity $\theta_0$) under function composition. This semigroup structure is induced from the semigroup structure of $\LL$.

\begin{example}
Fix $t_0 \in \RR$ and let $\TT := [t_0,\infty) \subseteq \RR$ be the time set with corresponding duration set $\DD$. The initial value problem consisting of the exponential growth model $f'(t) = f(t)$ and initial condition $(s,x) \in \TT \times \RR_{>0}$ has the solution $f \colon \TT \to \RR_{> 0}$ where $t \mapsto x e^{t-s}$. Since the exponential growth model depends only on the state of the system $f(t)$ and not explicitly on $t$, it generates an autonomous dynamical system. Define the state space $X := \RR_{>0}$ and the evolution function $\eta \colon \DD \times X \to X$ by $\eta((s,t),x) := xe^{t-s}$ for all $((s,t),x) \in \DD \times X$, which satisfies the properties of autonomy, identity, and the causal relation. Noting that $\LL = \RR_{\ge 0}$, define $\theta \colon \LL \times X \to X$ by $\theta (l,x) := \eta((s,t),x) = xe^{t-s} = xe^l$ for $(l,x) \in \LL \times X$ and $(s,t) \in \DD$ with $t-s = l$. Then the functions $(\theta_l)_{l \in \LL}$ on $X$ satisfy the semigroup property.
\end{example}

\subsubsection{Nonautonomous deterministic systems (process formalism)}
While autonomous systems are time invariant, nonautonomous systems are time varying, so the dynamics depend explicitly on time points due to external influences. Nonautonomous systems are particularly relevant for describing living organisms, which are thermodynamically open systems that continuously exchange energy (including matter) with their local environment: for example, the extracellular matrix influences cell behaviour, including the regulation of cell fate, through physical and biochemical interactions. The complexity of living organisms arises in part from the nonautonomy of their subsystems, limiting the extent to which studying subsystems in isolation is productive. Nonautonomous systems are more complicated than autonomous systems, both physically and mathematically. There are two main descriptions of nonautonomous deterministic dynamical systems, the \emph{process (two-parameter semigroup) formalism} and the \emph{skew-product flow formalism}.

The process formalism develops in a similar manner to the time-point formulation of autonomous systems, without the assumption of autonomy for the evolution function $\eta \colon \DD \times X \to X$. The dynamics may therefore depend on the initial and final time points.

\begin{example}\label{ex:process}
Fix $t_0 \in \RR_{>0}$ and let $\TT := [t_0,\infty) \subseteq \RR$ be the time set with corresponding duration set $\DD$. The initial value problem consisting of the model $f'(t) = (1/t) f(t)$ and initial condition $(s,x) \in \TT \times \RR_{>0}$ has the solution $f \colon \TT \to \RR_{> 0}$ where $t \mapsto (t/s) x$. Since the  model depends explicitly on the time points $t$ and $s$, it generates a nonautonomous dynamical system. We can view this system as a growing population that is subject to a disturbance that reduces the rate of growth more as time progresses. Define the state space $X := \RR_{>0}$ and the evolution function $\eta \colon \DD \times X \to X$ by $\eta((s,t),x) := (t/s) x$ for all $((s,t),x) \in \DD \times X$, which satisfies the identity property and the causal relation, however not autonomy.
\end{example}

\subsubsection{Nonautonomous deterministic systems (skew-product flow formalism)}
The notion of a \emph{skew-product flow} was originally developed within ergodic theory and topological dynamics. It forms the basis for a natural description of a nonautonomous dynamical system as a state space $X$ on which the dynamics are driven by an autonomous dynamical system, the \emph{driving system}, for which the evolution function $\theta \colon \LL \times P \to P$ (see Section~\ref{subsubsec:autonomous_duration}) acts on the \emph{base} or \emph{parameter space} $P$. To model the dynamics on $X$ we incorporate the driving mechanism provided by the autonomous system using a \emph{cocycle} function $\phi \colon \LL \times P \times X \to X$ whereby, given initial parameter $p \in P$ and initial state $x \in X$, the system transitions to the final state $\phi(l,p,x) \in X$ after the duration length $l \in \LL$. The following three properties characterize the skew-product flow formalism of a nonautonomous deterministic system.

\textbf{Determinism:} The cocycle $\phi$, which describes the state transitions, is a function so the system is deterministic.

\textbf{Identity:} For durations of length zero we specify no change in state, so $\phi(0,p,x) = x$ for all $p \in P$ and $x \in X$.

\textbf{Cocycle property:} The causal relation is provided by the cocycle property $\phi(s+t,p,x) = \phi(t,\theta(s,p),\phi(s,p,x))$ for all $s$, $t \in \LL$, $p \in P$, and $x \in X$, which also describes the influence of the driving system on the state space dynamics. The cocycle property ensures that the following two state transitions on $X$ are the same, with initial parameter $p \in P$, initial state $x \in X$, and duration lengths $s$, $t \in \LL$. First, the state transition from $x$ to $\phi(s+t,p,x)$ after duration length $s+t$. Second, the state transition from $x$ to $\phi(s,p,x)$ with corresponding parameter transition from $p$ to $\theta(s,p)$, after duration length $s$; followed by the state transition from $\phi(s,p,x)$ to $\phi(t,\theta(s,p),\phi(s,p,x))$ after duration $t$. The cocycle property is the nonautonomous analogue of the semigroup property of autonomous deterministic systems, and in fact generalizes the latter: just take the parameter space $P$ to be a singleton.

\begin{example}
Using the initial value problem in Example~\ref{ex:process}, we generate a nonautonomous system with the skew-product flow formalism. Define the state space $X := \RR_{>0}$, parameter space $P := \RR_{>0}$, and the cocycle function $\phi \colon \LL \times P \times X \to X$ such that $\phi(l,p,x) = ((l+p)/p) x$ for all $(l,p,x) \in \LL \times P \times X$. For the autonomous driving system define the evolution function $\theta \colon \LL \times P \to P$ by $\theta(l,p) := l + p$ for all $(l,p) \in \LL \times P$. Then the cocycle satisfies the identity and cocycle properties.
\end{example}

\subsubsection{Random systems}
While some intercellular and microenvironmental processes underlying the observed dynamics of cell fate are deterministic, other processes may have a degree of randomness, which produces time-varying dynamics that are unpredictable. \emph{Random systems} are a very natural framework for developing models of cell fate dynamics which incorporate randomness, and are developed in a similar manner to the skew-product flow formalism of nonautonomous deterministic systems, except the driving system now provides a model for the randomness.

\section{Towards a mathematical framework for modelling cell fate dynamics}\label{sec:Framework}
Here we provide a considered and formal approach towards establishing a general mathematical framework for developing models of cell fate dynamics. The aims of this framework are threefold: to guide the modelling process, particularly with respect to the assumptions appropriate for the physical systems; to ensure the models faithfully represent the physical systems; and, minimising replication of effort by classifying physical systems up to a suitable notion of equivalence.

Based on our motivational discussion in Section~\ref{sec:Represent}, the theory of random dynamical systems is a natural mathematical framework for developing models of cell fate dynamics. Moreover, it is apparent that any simpler framework would not adequately represent our experience of the complexity of cell fate dynamics. We therefore consider random dynamical systems in detail, beginning with autonomous and nonautonomous dynamical systems which form the basis for random systems. Since our aim in this paper is the consideration of a framework for model development, no particular models are discussed.

Randomness, broadly defined as unpredictability, arises within the cell--microenvironment system from two distinct origins: \emph{noise} and \emph{functional randomness}. Formally, noise is a disturbance to a system that negatively impacts its function. An observation of randomness within the cell--microenvironment system is often referred to as noise, implying a disturbance to cellular control or homeostasis. Of fundamental importance for understanding cell behaviour, however, is the recognition that an observation of `noise' may rather correspond to an operative mechanism for adaptivity, diversity, or flexibility, contrary to the notion of a disturbance~\cite{Bravi2015Unconventionality}. Randomness that corresponds to an operative mechanism is described as functional randomness. It is readily apparent that engagement with the dichotomy between functional randomness and noise is essential for understanding cell fate dynamics, and cell behaviour in general, as well as for developing models that are mechanistic rather than phenomenological.

\subsection{Autonomous deterministic dynamical systems}
An \emph{autonomous deterministic dynamical system} is a tuple $(X,\LL,\theta)$ where $X$ is the state space, the totally-ordered commutative semigroup $\LL$ is the time-parameter set, and the evolution function $\theta \colon \LL \times X \to X$ is a left semigroup action of $\LL$ on $X$ such that:
\begin{enumerate}
\item If $\LL$ has an identity element $e$ then $\theta(e,x) = x$ for all $ x \in X$.
\item (Semigroup property) $\theta(t+s,x) = \theta(t,\theta(s,x))$ for all $s$, $t \in \LL$ and $x \in X$.
\end{enumerate}
We may want to assume that $\LL$ has further properties, for example if $\LL$ is a non-monoid semigroup then we may assume that $\LL$ equals its positive cone $\LL^+$. We refer to $\LL$ as the \emph{time-parameter set}, rather than a time set or duration-length set, since it is now an abstract semigroup, and elements of $\LL$ are called \emph{time}. For $(s,x) \in \LL \times X$ the system describes the transition from the initial state $x \in X$ to the final state $\theta(s,x) \in X$ after time $s$.

We now consider a simple example of an autonomous deterministic dynamical system.

\begin{example}
Let $n \in \ZZ_{>0}$, let $X := [n]^{\ZZ}$ be the set of all bi-infinite sequences with elements in $[n]$, and let $\LL := \ZZ_{\ge0}$. Let $\sigma \colon X \to X$ be the left shift map such that $\sigma(x) (m) := x(m+1)$ for all $x \in X$ and $m \in \ZZ$. Define the evolution function $\theta \colon \LL \times X \to X$ such that $\theta(m,x) := \sigma^m(x)$ for $(m,x) \in \LL \times X$. Then $(X,\LL,\theta)$ is an autonomous deterministic dynamical system.
\end{example}

The definition of an autonomous deterministic dynamical system is abstract, and in particular the state space $X$ is just a set. In applications we often require the addition of some structure to $X$, and accordingly dynamical systems theory encompasses three main (overlapping) subareas~\cite{Feng2007Networks}: \emph{topological dynamics} is concerned with the study of \emph{topological dynamical systems}, informally the actions of continuous functions on topological spaces; \emph{differentiable dynamics} is concerned with the study of \emph{differentiable dynamical systems}, informally the actions of differentiable functions on manifolds; and, \emph{ergodic theory} is concerned with the study of \emph{measure-preserving dynamical systems}, informally the measure-preserving actions of measurable functions on measure spaces. When we add structure to the state space we require that the evolution function preserves this structure in an appropriate manner. We discuss topological and, especially, measure-preserving dynamical systems in more detail since they are immediately relevant.

\textbf{Topological dynamical systems:} A \emph{topological dynamical system} is a dynamical system $((X,\tau),(\LL,+,\T),\theta)$ where $(X,\tau)$ is a topological space, $(\LL,+,\T)$ is a topological semigroup, and $\theta \colon \LL \times X \to X$ is a jointly continuous function. It is usually assumed that $X$ has additional topological properties such as being Hausdorff, metrizable, locally compact, or compact. Note that we can write a topological dynamical system $((X,\tau),(\LL,+,\T),\theta)$ in the standard dynamical system form $(X,\LL,\theta)$ with the remaining information implied.

\textbf{Measurable dynamical systems:} A \emph{measurable dynamical system}~\cite[Appendix A, page 536]{Arnold1998Random} is a tuple $((X,\X),(\LL,+,\Ll),\theta)$ where:
\begin{enumerate}
    \item $(X,\X)$ is a measurable space, so $\X$ is a $\sigma$-algebra over the state space $X$.
    \item $\LL$ is the time-parameter set where $(\LL,+,\Ll)$ is a measurable semigroup.
    \item The evolution function $\theta \colon \LL \times X \to X$ is $(\Ll \otimes \X,\X)$-measurable.
    \item If $\LL$ has an identity $e$ then $\theta(e,x) = x$ for all $x \in X$.
    \item (Semigroup property) $\theta(t,\theta(s,x)) = \theta (t + s,x)$ for all $s$, $t \in \LL$ and $x \in X$.
\end{enumerate}   
Note that we can write a measurable dynamical system $((X,\X),(\LL,+,\Ll),\theta)$ in the standard dynamical system form $(X,\LL,\theta)$ with the remaining information implied.

\textbf{Measure-preserving dynamical systems:} Recall that a probability space $(X,\X,\PP)$ is a measure space with total measure one, that is $\PP(X) = 1$, where the set $X$ is the \emph{sample space} of all possible \emph{outcomes}, the $\sigma$-algebra $\X$ on $X$ is the \emph{event space}, and the \emph{probability measure} $\PP \colon \X \to [0,1]$ is a nonnegative $\sigma$-additive measure. A \emph{measure-preserving dynamical system}, or \emph{metric dynamical system}, is a tuple $((X,\X,\PP),(\LL,+,\Ll),\theta)$ where:
\begin{enumerate}
    \item $((X,\X),(\LL,+,\Ll),\theta)$ is a measurable dynamical system.
    \item $(X,\X,\mathbb{P})$ is a probability space.
    \item For each $t \in \LL$ the map $\theta_t \colon X \to X$, where $\theta_t(x) := \theta(t,x)$ for all $x \in X$, is measurable and preserves the measure $\PP$, that is, $\PP(\theta_t^{-1}(A)) = \PP(A)$ for all $A \in \X$ where $\theta_t^{-1}(A)$ is the preimage of $A$ under $\theta_t$. $\mathbb{P}$ is said to be an \emph{invariant} measure for $\theta_t$.
\end{enumerate}
We can write a measure-preserving dynamical system $((X,\X,\PP),(\LL,+,\Ll),\theta)$ in the standard dynamical system form $(X,\LL,\theta)$ with the remaining information implied.

Measure-preserving dynamical systems are the driving systems of random dynamical systems, so we consider a standard example.

\begin{example}[Bernoulli scheme]\label{ex:Bernoulli}\leavevmode

Let $n \in \ZZ_{>0}$ and let $(p_i)_{i=1}^n$ be a positive probability vector. Define the set function $\mu \colon 2^{[n]} \to [0,1]$ by $\mu(M) := \sum_{i \in M} p_i$ for $M \in 2^{[n]}$, where the empty sum evaluates to zero. Then $([n],2^{[n]},\mu)$ is a probability space.

Let $\Omega := [n]^{\ZZ}$ be the set of all bi-infinite sequences with elements in $[n]$. Let $\PS$ be the countable direct product $\bigotimes_{-\infty}^{\infty} 2^{[n]}$, which has a basis consisting of the cylinder sets of $\Omega$: for $z \in \ZZ$, $r \in \ZZ_{>0}$, and $(a_i)_{i=1}^r \in [n]^{[r]}$ define the cylinder set $C_z[(a_i)_{i=1}^r] := \{\, \omega \in \Omega \mid \text{$\omega_z = a_1, \omega_{z+1} = a_2, \ldots, \omega_{z+r-1} = a_r$}\,\}$. $\PS$ is called the cylindrical $\sigma$-algebra generated by the cylinder sets in $\Omega$. Let $\rho$ be the product measure on $\PS$, which satisfies $\rho(C_z[(a_i)_{i=1}^r]) = \prod_{i=1}^r p_{a_i}$ for a cylinder set $C_z[(a_i)_{i=1}^r]$. Then $(\Omega,\PS,\rho)$ is a probability space.

The left shift map $\sigma \colon \Omega \to \Omega$ such that, for $\omega \in \Omega$, $\sigma(\omega) (m) := \omega(m+1)$ for $m \in \ZZ$ is a measurable and measure preserving bijection. The inverse of $\sigma$ is the right shift map $\sigma^{-1} \colon \Omega \to \Omega$ where, for $\omega \in \Omega$ and $m \in \ZZ$, $\sigma^{-1}(\omega) (m) := \omega(m-1)$, which is measurable and measure preserving. Define the evolution function $\theta \colon \ZZ \times \Omega \to \Omega$ such that $\theta(z,\omega) := \sigma^z(\omega)$ for $(z,\omega) \in \ZZ \times \Omega$, noting that the $\sigma$-algebra on $\ZZ$ is the power set $2^{\ZZ}$. $\theta$ is $(2^{\ZZ} \otimes \PS, \PS)$-measurable, since if $C_z[(a_i)_{i=1}^r]$ is a cylinder set in $\Omega$, for some $z \in \ZZ$ and $r \in \ZZ_{>0}$, then $\theta^{-1}(C_z[(a_i)_{i=1}^r]) = \{\, (t,\omega) \in \ZZ \times \Omega \mid \theta(t,\omega) \in C_z[(a_i)_{i=1}^r] \,\} = \bigcup_{k \in \ZZ} \{k\} \times C_{z+k}[(a_i)_{i=1}^r] \in 2^{\ZZ} \otimes \PS$. Further, $\theta(0,\omega) = \omega$ for all $\omega \in \Omega$, and the semigroup property holds since $\theta(t+s,\omega) = \sigma^{t+s}(\omega) = \sigma^t(\sigma^s(\omega)) = \theta(t,\theta(s,\omega))$ for all $s$, $t \in \ZZ$ and $\omega \in \Omega$. So $((\Omega,\PS),(\ZZ,+,2^{\ZZ}),\theta)$ is a measurable dynamical system.

For $t \in \ZZ$ the map $\theta_t \colon \Omega \to \Omega$ is measurable since $\theta_t = \sigma^t$ is the composition of the measurable function $\sigma$ on $\Omega$. Further, $\theta_t$ preserves the measure $\rho$ since if $C_z[(a_i)_{i=1}^r]$ is a cylinder set in $\Omega$ for some $z \in \ZZ$ and $r \in \ZZ_{>0}$ then $\theta_t^{-1}(C_z[(a_i)_{i=1}^r]) = C_{z+t}[(a_i)_{i=1}^r]$, and the result follows. It follows that the tuple $((\Omega,\PS,\rho),(\ZZ,+,2^{\ZZ}),\theta)$ is a metric dynamical system referred to as a two-sided Bernoulli scheme with probability vector $(p_i)_{i=1}^n$.

A one-sided Bernoulli scheme with probability vector $(p_i)_{i=1}^n$ is constructed similarly, but with time set $\ZZ_{\ge0}$ instead of $\ZZ$.
\end{example}

Autonomous dynamical systems can be generated by deriving the state space and evolution function from autonomous differential equations, which do not depend explicitly on the independent variable (generally time). Alternatively, the state space and evolution function can be specified directly, as in Example~\ref{ex:Bernoulli}.

\subsection{Nonautonomous deterministic dynamical systems}
We discuss nonautonomous dynamical systems since they are highly relevant for modelling biological systems, and also motivate the definition of a random dynamical system. A \emph{nonautonomous deterministic dynamical system} can be defined with either the \emph{process} (or \emph{two-parameter semigroup}) formalism or the \emph{skew product flow} formalism~\cite{Kloeden2011Nonautonomous}. Each of these two formalisms has certain advantages over the other, depending on the specific application.

\subsubsection{Process formalism}
A \emph{nonautonomous deterministic dynamical system (process)} is a tuple $(X,\DD,\phi)$ where $X$ is the state space, $\DD := \TT^{(2,\le)}$ for some time-parameter set $\TT$ where $(\TT,+,\le)$ is a totally-ordered commutative semigroup, making $\DD$ a commutative semigroup under the coordinate-wise addition, and the \emph{process} $\phi \colon \DD \times X \to X$ is a left semigroup action of $\DD$ on $X$ such that:
\begin{enumerate}
    \item (Initial value) $\phi((t,t),x) = x$ for all $t \in \TT$ and $x \in X$.
    \item (Causality) $\phi((s,t),x) = \phi((u,t),\phi((s,u),x))$ for all $s$, $u$, $t \in \TT$ with $s \le u \le t$.
\end{enumerate}

\subsubsection{Skew product flow formalism}
In this formalism, a \emph{cocycle} describes the evolution of the nonautonomous dynamical system with respect to a driving autonomous dynamical system. A similar formalism is employed in random dynamical systems.

A \emph{nonautonomous deterministic dynamical system (skew-product flow)} is a tuple $(X,\phi,(P,\LL,\theta))$ where $X$ is the state space (or fibre), $(P,\LL,\theta)$ is an autonomous dynamical system, where we call $P$ the \emph{parameter space}, and $\phi \colon \LL \times P \times X \to X$ is a \emph{cocycle} on $X$ with \emph{base} or \emph{driving system} $(P,\LL,\theta)$ such that:
\begin{enumerate}
    \item If $\LL$ has an identity element $e$ then $\phi(e,p,x) = x$ for all $(p,x) \in P \times X$.
    \item (Cocycle property) $\phi(t+s,p,x) = \phi(t,\theta(s,p),\phi(s,p,x))$ for all $s$, $t \in \LL$ and $(p,x) \in P \times X$.
\end{enumerate}
The cocycle property generalizes the semigroup property of autonomous systems. We have defined the cocycle to allow for both forward and backward time, however sometimes it is only defined in forward time. In this case, the domain of $\phi$ is $\LL^+ \times P \times X \to X$ where $\LL^+$ is the positive (or nonnegative) cone of $\LL$.

The \emph{skew product flow} associated with the nonautonomous system is the tuple $(P \times X,\LL,\eta)$, which is an autonomous dynamical system where $\eta \colon \LL \times (P \times X) \to P \times X$ is given by $\eta(t,(p,x)) := (\theta(t,p),\phi(t,p,x))$. Further, note that a system with the process formalism can be interpreted as a nonautonomous system with the skew product flow formalism~\cite[see Example 2.8, page 29, Chapter 2]{Kloeden2011Nonautonomous}.

\subsection{Random dynamical systems}
A \emph{random dynamical system}~\cite{Arnold1998Random} consists of two main components: a \emph{metric dynamical system} for the \emph{base flow}, which provides a model for the randomness; and a model of the system dynamics on the state space which are influenced by the randomness, described as a cocycle over the base flow. A random dynamical system is, therefore, a type of nonautonomous system.

A \emph{random dynamical system} is a tuple $((X,\X),\phi,((\Omega,\F,\PP),(\LL,+,\Ll),\theta))$ where:
\begin{enumerate}
     \item $(X,\X)$ is a measurable space, with $X$ the \emph{state space}.
    \item $((\Omega,\F,\PP),(\LL,+,\Ll),\theta)$ is a metric dynamical system, with $\Omega$ the \emph{parameter space}.
    \item $\phi \colon \LL \times \Omega \times X \to X$ is a cocycle on $X$ driven by $((\Omega,\F,\PP),(\LL,+,\Ll),\theta)$, that is:
         \begin{enumerate}
            \item If $\LL$ has an identity element $e$ then $\phi(e,\omega,x) = x$ for all $\omega \in \Omega$ and $x \in X$.
            \item (Cocycle property) $\phi(t + s,\omega,x) = \phi(t,\theta(s,\omega),\phi(s,\omega,x))$ for all $s$, $t \in \LL$, $\omega \in \Omega$, and $x \in X$.
        \end{enumerate}
    \item $\phi$ is $(\Ll \otimes \F \otimes \X, \X)$-measurable.
\end{enumerate}

Here we have assumed that $\phi$ is a \emph{perfect cocycle}, whereby the cocycle property holds identically. If the cocycle property holds for fixed $s$ and for all $t \in \LL$, $\PP$-almost surely where the exceptional set may depend on $s$, then $\phi$ is a \emph{crude cocycle}. If the cocycle property holds for fixed $s$, $t \in \LL$, $\PP$-almost surely where the exceptional set may depend on both $s$ and $t$, then $\phi$ is a \emph{very crude cocycle}.

We can add further structure to the state space $X$. For example, a random dynamical system is \emph{continuous} or \emph{topological} when $(X,\tau)$ is a topological space, $(\LL,+,\T)$ is a topological semigroup, and the function $\phi_{\omega} \colon \LL \times X \to X$ given by $\phi_{\omega}(t,x) := \phi(t,\omega,x)$ is jointly continuous for all $(t,x) \in \LL \times X$ and $\omega \in \Omega$, where $\Omega$ is not necessarily a topological space. Note that while joint continuity in $t$ and $x$ is often assumed, and holds for many examples of random dynamical systems, the theory of random attractors requires only continuity in $x$, that is, the function $\phi_{(t,\omega)} \colon X \to X$ given by $\phi_{(t,\omega)}(x) := \phi(t,\omega,x)$ is continuous for all $(t,\omega) \in \LL \times \Omega$ and $x \in X$.

\begin{figure}[ht]
\centering\includegraphics[width=1\textwidth]{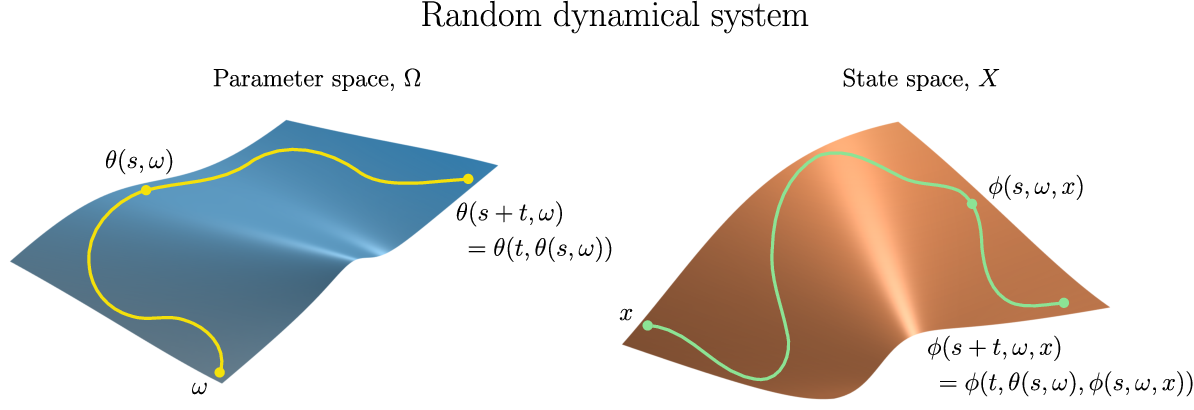}
\caption{Schematic of a random dynamical system $((X,\X),\phi,((\Omega,\F,\PP),(\LL,+,\Ll),\theta))$, where $\Omega$ is the parameter space with evolution function $\theta$, and $X$ is the state space with cocycle $\phi$. Note that times $s$, $t \in \LL$ satisfy $t > s$.}
\label{fig:RDS}
\end{figure}

Random dynamical systems can be generated by stochastic or random differential equations, or constructed directly as in Example~\ref{Ex:RDS}. We schematically illustrate the concept of a random dynamical system in Figure~\ref{fig:RDS}.

\begin{example}[Random systems generalize deterministic measurable systems]\leavevmode

Suppose that $((X,\X),\phi,((\Omega,\F,\PP),(\LL,+,\Ll),\theta))$ is a random dynamical system where the cocycle $\phi$ does not depend on the parameter space $\Omega$, for example when $\Omega$ is a singleton set. Define $\widehat{\phi} \colon \LL \times X \to X$ such that $\widehat{\phi}(t,x) := \phi(t,\omega,x)$ for $(t,x) \in \LL \times X$ and any $\omega \in \Omega$, which is well defined since $\phi$ does not depend on $\Omega$. Then $((X,\X),(\LL,+,\Ll),\widehat{\phi})$ is a deterministic measurable dynamical system. Conversely, every deterministic measurable dynamical system is a random dynamical system where the cocycle does not depend on the parameter space $\Omega$.
\end{example}

\begin{example}[A continuous random dynamical system driven by a Bernoulli scheme]\label{Ex:RDS}\leavevmode

Let $(X,\B(X))$ be the measurable space where $X := [0,1]$ has the standard topology and $\B(X)$ is the Borel $\sigma$-algebra over $X$. Let $((\Omega,\PS,\rho),(\ZZ_{>0},+,2^{\ZZ_{>0}}),\theta)$ be the one-sided Bernoulli scheme with positive probability vector $(p_i)_{i=1}^n$, for some $n \in \ZZ_{>0}$, as described in Example~\ref{ex:Bernoulli}.

Suppose that $(f_i)_{i=1}^n$ is a finite sequence of continuous transformations on $X$. Since each $f_i$, $i \in [n]$, is continuous it is $(\B(X),\B(X))$-measurable. Define the function $\phi \colon \ZZ_{>0} \times \Omega \times X \to X$ by $\phi(z,\omega,x) := f_{\omega(z-1)} \circ \cdots \circ f_{\omega(0)} (x)$ for $(z,\omega,x) \in \ZZ_{>0} \times \Omega \times X$, with the convention that $\phi(0,\omega,x) = x$ for all $(\omega,x) \in \Omega \times X$. To see that $\phi$ has the cocycle property, let $s$, $t \in \ZZ_{>0}$, $\omega \in \Omega$, and $x \in X$, then
\begin{align*}
\hspace{-0.8cm}\phi(t+s,\omega,x)
&= f_{\omega(t+s-1)} \circ \cdots \circ f_{\omega(0)} (x) = f_{\theta(s,\omega)(t-1)} \circ \cdots \circ f_{\theta(s,\omega)(0)} \circ f_{\omega(s-1)} \circ \cdots \circ f_{\omega(0)} (x)\\[0.2cm]
&= \phi(t,\theta(s,\omega),f_{\omega(s-1)} \circ \cdots \circ f_{\omega(0)} (x)) = \phi(t,\theta(s,\omega),\phi(s,\omega,x)),
\end{align*} 
as required. Further, $\phi$ is $(2^{\ZZ_{>0}} \otimes \PS \otimes \B(X), \B(X))$-measurable, since if $S \in \B(X)$ then $\phi^{-1}(S) = \{\, (z,\omega,x) \mid \phi(z,\omega,x) \in S \,\} = \bigcup_{z \in \ZZ_{>0}, \, \omega \in \Omega} \{z\} \times C_0[(\omega(i)i)_{i=0}^{z-1}] \times (f_{\omega(z-1)} \circ \cdots \circ f_{\omega(0)})^{-1}(S) \in 2^{\ZZ_{>0}} \otimes \PS \otimes \B(X)$.

Note that $X = [0,1]$ is a topological space with the standard topology, and $\ZZ_{>0}$ is a discrete topological space. So, for $\omega \in \Omega$, the function $\phi_{\omega} \colon \ZZ_{>0} \times X \to X$ given by $\phi_{\omega}(t,x) := \phi(t,\omega,x)$ satisfies, for an open set $U$ in $X$,
\begin{align*}
\phi_{\omega}^{-1}(U) &= \{\, (t,x) \in \ZZ_{>0} \times X \mid f_{\omega(t-1)} \circ \cdots \circ f_{\omega(0)} (x) \in U\,\}\\[0.2cm]
&= \bigcup_{v \in \ZZ_{>0}} \{v\} \times (f_{\omega(z-1)} \circ \cdots \circ f_{\omega(0)})^{-1}(U),
\end{align*}
which is an open set in the product space $\ZZ_{>0} \times X$, hence $\phi_{\omega}$ is a continuous function. It follows that the tuple $((X,\B(X)),\phi,((\Omega,\PS,\rho),(\ZZ_{>0},+,2^{\ZZ_{>0}}),\theta))$ is a continuous random dynamical system driven by the one-sided Bernoulli scheme $((\Omega,\PS,\rho),(\ZZ_{>0},+,2^{\ZZ_{>0}}),\theta)$ with probability vector $(p_i)_{i=1}^n$.
\end{example}

\subsection{Asymptotic and long transient dynamics}

\subsubsection{Asymptotic dynamics}
Time-asymptotic, or simply asymptotic, dynamics refers to the long-time behaviour of a dynamical system, and may involve notions such as omega-limit sets, invariant sets, or attractors. Here, we focus on attractors, as they are often discussed within the context of cell fate. The concept of an \emph{attractor}, while evocative, is in general quite complicated: multiple definitions of attractors are found in the literature~\cite{Milnor1985,Feng2007Networks}, however providing precise definitions continues to be a challenge in various situations; and, dynamical systems may have no attractors or may have multiple, possibly infinitely many, attractors.

The notion of an attractor is based on a suitable definition of closeness of the system states, leading to different definitions of attractors: for example, topological, metric, measure, and statistical. There are two general properties that we may assume of an attractor: it must describe the asymptotic behaviour of a large set of initial
conditions, called the \emph{basin (or realm) of attraction}; and all parts of the attractor should be involved in describing the asymptotic behaviour of the points in the basin of attraction.

Here we describe some key examples of attractors for autonomous, nonautonomous, and random dynamical systems~\cite{Kloeden2011Nonautonomous}. For autonomous dynamical systems, limiting objects such as attractors are independent of time. For nonautonomous systems, however, the situation may be much more complicated: while the limiting objects for autonomous systems are still applicable, they often exclude important dynamics, so we need to consider limiting objects that are time dependent. Definitions of attractors for nonautonomous systems must, therefore, account for the dynamic nature of limiting objects. Two such notions of attractors are the \emph{forward attractor}, which involves a dynamic limiting object moving forward in time, and the \emph{pullback attractor}, which involves a limiting object that is fixed in time and with progressively earlier starting times. Forward (resp. pullback) attraction is based on future (resp. past) states of the system. Forward and pullback attractors are generally distinct objects, but are equivalent when the system is autonomous.

For simplicity we assume that the state space $X$ is the underlying set of a complete metric space $(X,d_X)$. Recall that a subset $A \subseteq X$ is \emph{compact} when every open cover of $A$ has a finite subcover. For nonempty subsets $V$ and $W$ of $X$ denote by $d_{\Hh}(V,W) := \sup_{v\in V} \inf_{w\in W} d_X(v,w)$ the \emph{directed (nonsymmetric) Hausdorff distance} from $V$ to $W$. Denote by $\TT \in \{\ZZ,\RR\}$ the time set.

\textbf{Autonomous:} Let $(X,\TT_{\ge 0},\theta)$ be an autonomous deterministic dynamical system where the evolution function $\theta \colon \TT_{\ge 0} \times X \to X$ is continuous. A \emph{global attractor} of $(X,\TT_{\ge 0},\theta)$ is a nonempty compact subset $A \subseteq X$ such that: $A$ is strictly invariant under $\theta$, so $\theta(t,A) = A$ for all $t \in \TT_{\ge 0}$; and, $A$ attracts all nonempty bounded subsets $B \subseteq X$, so $\lim_{t \to \infty} d_{\Hh}(\theta(t,B),A) = 0$. Note that a global attractor is unique, and contains all of the asymptotic dynamics of the system. If we are only interested in the asymptotic behaviour, rather than the transient behaviour, then the (compact) global attractor $A$ generally provides a dimensional reduction of the state space $X$.

\textbf{Nonautonomous (process):} Let $(X,\DD,\phi)$ be a nonautonomous deterministic dynamical system where the semigroup $\DD := \TT^{(2,\le)}$ is the duration set, and the process $\phi \colon \DD \times X \to X$ is continuous. A subset $\N$ of the extended phase space $\TT \times X$ is a \emph{nonautonomous set} of the process $\phi$, and for every $t \in \TT$ the \emph{$t$-fibre} of $\N$ is the set $N_t := \{\,x \in X \mid (t,x) \in \N \,\}$. A nonautonomous set $\N$ is \emph{invariant} if $\phi(s,t,N_s) = N_t$ for all $(s,t) \in \DD$. For a given topological property, such as compactness or closedness, we say that $\N$ has the property if and only if each fibre of $\N$ has the property.

Let $\A$ be a nonempty, compact, and invariant nonautonomous set of $\phi$. Then $\A$ is a \emph{forward attractor} if $\lim_{t \to \infty} d_{\Hh}(\phi(s,t,B),A_t) = 0$ for initial time $s \in \TT$ and for all bounded subsets $B \subseteq X$. Further, $\A$ is a \emph{pullback attractor} if $\lim_{s \to -\infty} d_{\Hh}(\phi(s,t,B),A_t) = 0$ for initial time $s \in \TT$ and for all bounded subsets $B \subseteq X$.

Informally, if $\A$ is a forward attractor then each $A_t$ is a cross section of $\A$ at time $t \in \TT$. For a fixed start time $s \in \TT$, $\A$ forward attracts every bounded subset of $X$ in the sense that, for any bounded $B \subseteq X$, each orbit starting in $B$ at time $s$ will approach $A_t$ as $t \to \infty$: $A_t$ is like a target moving forward in time. Similarly, if $\A$ is a pullback attractor then, for a fixed time $t \in \TT$, $\A$ pullback attracts every bounded subset of $X$ in the sense that, 
for any bounded $B \subseteq X$, each orbit starting in $B$ at time $s$ will become arbitrarily close to $A_t$ at time $t$ as $s \to -\infty$: $A_t$ is like a fixed target.

\textbf{Nonautonomous (skew-product flow):} Let $(X,\phi,(P,\TT,\theta))$ be a nonautonomous deterministic dynamical system where $\phi$ is a cocycle on $X$ driven by the autonomous system $(P,\TT,\theta)$ with $(P,d_P)$ a metric space. A subset $\N$ of the extended phase space $P \times X$ is a \emph{nonautonomous set} of the skew product flow $(\phi,\theta)$, and for every $p \in P$ the \emph{$p$-fibre} of $\N$ is the set $N_p := \{\,x \in X \mid (p,x) \in \N \,\}$. A nonautonomous set $\N$ is \emph{invariant} if $\phi(t,p,N_p) = N_{\theta(t,p)}$ for all $t \ge 0$ and $p \in P$. For a given topological property, such as compactness or closedness, we say that $\N$ has the property if and only if each fibre of $\N$ has the property.

Let $\A$ be a nonempty, compact, and invariant nonautonomous set of $\phi$. Then $\A$ is a \emph{forward attractor} if the \emph{forward convergence} $\lim_{t \to \infty} d_{\Hh}(\phi(t,p,B),A_{\theta(t,p)}) = 0$ holds for all nonempty bounded subsets $B \subseteq X$ and $p \in P$. Further, $\A$ is a \emph{pullback attractor} if the \emph{pullback convergence} $\lim_{t \to \infty} d_{\Hh}(\phi(t,\theta(-t,p),B),A_p) = 0$ holds for all nonempty bounded subsets $B \subseteq X$ and $p \in P$.

\textbf{Random:} Let $((X,\mathcal{A}),\phi,((\Omega,\mathcal{F},\mathbb{P}),(\mathbb{T},+,\mathcal{T}),\theta))$ be a random dynamical system on the measurable space $(X,\mathcal{A})$ where $\A$ is the Borel $\sigma$-algebra on $X$, over the metric dynamical system $((\Omega,\mathcal{F},\mathbb{P}),(\mathbb{T},+,\mathcal{T}),\theta))$, with time set $\mathbb{T}$ and cocycle $\phi$ over $\theta$. We assume further that the metric space $(X,d_X)$ is separable.

Attractors of random dynamical systems are similar to those of deterministic nonautonomous systems (skew-product flow), though have characteristic properties with respect to measurability which are described with the notion of a random set. A map $K \colon \Omega \to 2^X$ is a \emph{random set} if its graph $\Gamma(K) := \{\,(\omega,x) \in \Omega \times X \mid x \in K(\omega)\,\}$ is measurable, that is, $\Gamma(K) \in \F \otimes \A$. We henceforth assume that a given random set is nonempty $\PP$-almost surely.

A map $K \colon \Omega \to 2^X$ is a \emph{closed (resp. compact) random set} if the sets $K(\omega)$ are closed (resp. compact) subsets of $X$ for all $\omega \in \Omega$, and if the map $\omega \mapsto d_{\Hh}(x,K(\omega))$ from $\Omega$ to $[0,\infty)$ is $(\F,\B([0,\infty)))$-measurable for all $x \in X$. Note that a closed random set is always a random set: the function $\pi \colon \Omega \times X \to [0,\infty)$ where $\pi(\omega,x) = d_{\Hh}(x,K(\omega))$ is jointly measurable by~\cite[Lemma 1.1, Page 2, Chapter 1]{Crauel2002Random}, since $\omega \mapsto \pi(\omega,x)$ is measurable for each $x \in X$, and $x \mapsto \pi(\omega,x)$ is continuous for each $\omega \in \Omega$ as a distance to a set function; since $\{0\} \in \B([0,\infty))$ it follows that $\Gamma(K) = \{\,(\omega,x) \in \Omega \times X \mid d_{\Hh}(x,K(\omega)) = 0\,\} = \pi^{-1}(\{0\}) \in \F \otimes \A$, where the first equality holds since each $K(\omega)$ is closed. Further, since compact sets are closed in metric spaces, a compact random set is also a random set.

Let $A \colon \Omega \to 2^X$ be a compact random set. $A$ is \emph{strictly $\phi$-invariant} if $\phi(t,\omega,A(\omega)) = A(\theta(t,\omega))$ $\PP$-almost surely for all $\omega \in \Omega$ and $t \ge 0$. Let $\D$ be a family of random sets $D \colon \Omega \to 2^X$. Then $A$ is a \emph{random forward attractor} for $\D$ if $A$ is strictly $\phi$-invariant, and $A$ is a \emph{forward attracting set} for $\D$ in the sense that $\lim_{t \to \infty} d_{\Hh}(\phi(t,\omega,D(\omega)),A(\theta(t,\omega))) = 0$, $\PP$-almost surely, for all $D \in \D$. Similarly, $A$ is a \emph{random pullback attractor} for $\D$ if $A$ is strictly $\phi$-invariant, and $A$ is a \emph{pullback attracting set} for $\D$ in the sense that $\lim_{t \to \infty} d_{\Hh}(\phi(t,\theta(-t,\omega),D(\theta(-t,\omega))),A(\omega)) = 0$, $\PP$-almost surely, for all $D \in \D$.

\begin{figure}[ht]
\centering\includegraphics[width=1\textwidth]{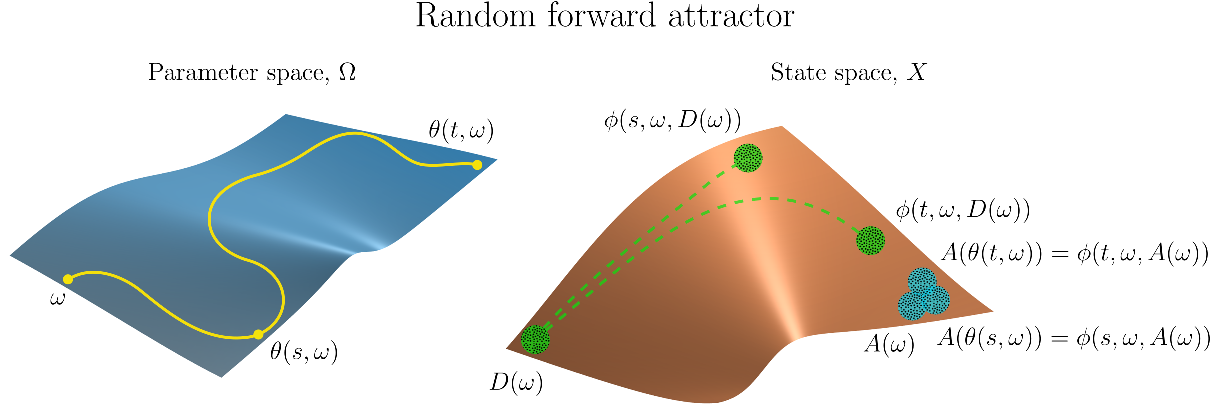}
\caption{Schematic of a random forward attractor for a random dynamical system $((X,\X),\phi,((\Omega,\F,\PP),(\LL,+,\Ll),\theta))$, where $\Omega$ is the parameter space with evolution function $\theta$, $X$ is the state space with cocycle $\phi$, and the compact random set $A \colon \Omega \to 2^X$ is a random forward attractor for the family $\D$ of random sets $D \colon \Omega \to 2^X$. The coloured discs with black dots represent the fibres, which are subsets of $X$, of the corresponding random sets. Note that the strict $\phi$-invariance of $A$ and the forward attraction of $\D$ towards $A$ hold $\PP$-almost surely, and times $s$, $t \in \LL$ satisfy $t > s$. }
\label{fig:Forward}
\end{figure}

\begin{figure}[ht]
\centering\includegraphics[width=1\textwidth]{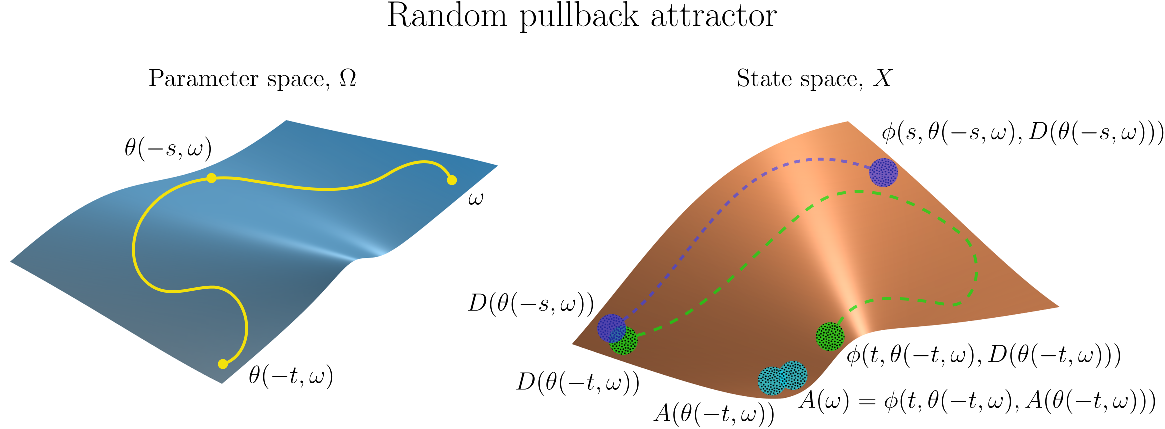}
\caption{Schematic of a random pullback attractor for a random dynamical system $((X,\X),\phi,((\Omega,\F,\PP),(\LL,+,\Ll),\theta))$, where $\Omega$ is the parameter space with evolution function $\theta$, $X$ is the state space with cocycle $\phi$, and the compact random set $A \colon \Omega \to 2^X$ is a random pullback attractor for the family $\D$ of random sets $D \colon \Omega \to 2^X$. The coloured discs with black dots represent the fibres, which are subsets of $X$, of the corresponding random sets. Note that the strict $\phi$-invariance of $A$ and the pullback attraction of $\D$ towards $A$ hold $\PP$-almost surely, and times $s$, $t \in \LL$ satisfy $t > s$. 
}
\label{fig:Pullback}
\end{figure}

We schematically illustrate the concepts of a random forward attractor and a random pullback attractor in Figures~\ref{fig:Forward} and \ref{fig:Pullback}, respectively.

\subsubsection{Long transient dynamics}
Cell fate is generally regarded as a dynamically asymptotic process, whereby cells develop into persistent cell types over a relatively short time. From the perspective of dynamical systems, it is usually assumed that the persistence of the cell type corresponds to an invariant region of state space, such as an attractor. The appeal of such a simple explanation, however, may result in an alternative interpretation being overlooked, namely \emph{long transient dynamics}~\cite{Liu2022Quantification}. Transient dynamics can persist for a very long time, can be quasi-stable, and may correspond to a non-asymptotic process. Examples of biological systems in which long transient dynamics are observed range from cell biology to ecology~\cite{Hastings2018Transient,Petrov2021}.

The existence of randomness in the cell--microenvironment system may suggest that cell fate dynamics involve long transients rather than attractors. In particular, randomness may prevent the system from residing near an attractor, or may prevent the existence of an attractor, thereby producing transient dynamics~\cite{Morozov2020}. It is increasingly apparent that long transient dynamics play a fundamental role in biological systems, and the structural aspects of dynamical systems that produce these behaviours are being identified~\cite{Koch2024Ghost}. Notably, long transients are likely to occur in systems with high dimensionality, randomness, or multiple time scales, all of which are present in cell systems~\cite{Hastings2018Transient}. The identification of long transients in cell systems would be essential for understanding cell fate dynamics. The subject of long transient dynamics is a relatively recent development, and while some major theoretical advances have been made, much more awaits.

\section{Conclusion}\label{sec:Conclusion}
Defining the notion of a cell type, and by extension cell state and cell fate, in a way that is both empirically-driven and precise is not an easy task~\cite{Dance2024What}. Perhaps the diversity of definitions, each tailored to specific contexts~\cite{Clevers2017What}, is a necessary consequence, and arguably reflects the intrinsically high-dimensional nature of cellular dynamics over the course of development. Yet, efforts to engage with this complexity either conceptually or mathematically have been limited. This is surprising in light of the existence of a rich and powerful theory of dynamical systems, only a fraction of which has made its way into our current understanding of cellular dynamics.

Part of the reason for this limited uptake may be due to the very idea that is meant to help our conceptual thinking about cell fate, the visual metaphor of an epigenetic landscape due to Waddington. While based partly on scientific observation, the two-dimensional landscape embedded in a three-dimensional state space is only a simplification of the very high-dimensional cellular state space, as clearly noted by Waddington himself~\cite[p. 31]{Waddington1957Strategy}. Further, the metaphor explicitly neglects the effect of the environment on phenotypes~\cite[p. 158]{Waddington1957Strategy}. We therefore need to take care with any literal interpretation of the landscape. Waddington was clear about this, unambiguously writing that ``the epigenetic landscape [...] cannot be interpreted rigorously''~\cite[p. 30]{Waddington1957Strategy}.

In this article we consider the process of cell fate specification from within a mathematical framework based on random dynamical systems. Since a cell is a thermodynamically open system, the appropriate state space for modelling cell fate is the cell--microenvironment system, which consists of the cell coupled with the microscale extracellular environment with which the cell interacts. Many processes within the cell--microenvironment system are subject to randomness, so modelling the cellular dynamics on the state space requires the random dynamical system formalism, whereby a metric dynamical system (a model for the randomness) drives the dynamics.

Random dynamical systems provide a natural framework for faithfully representing the characteristics of cell fate dynamics. Since the cell--microenvironment system is nonautonomous, a transition in the cell's state depends not just on the initial state but also on the time, so notions of stability for the system are much more complicated than for autonomous (time independent) dynamical systems. For example, while attractors for autonomous system are independent of time, attractors for random systems can be time dependent, and hence more complicated, such as the random pullback attractor. Engaging with the time-dependent nature of cell behaviour, however, is essential for understanding the intricacies of cellular dynamics. Indeed, the relevant attractors corresponding to the different possible fates of a pluripotent stem cell are arguably random: the dynamics incorporate an element of unpredictability due to the influence of randomness, making them explicitly depend on time and, therefore, effectively context-dependent.

The perspective afforded by random dynamical systems theory makes clear the inaccuracies associated with a literal interpretation of Waddington's epigenetic landscape. The landscape is typically modelled as a deterministic system, even though randomness is known to profoundly alter the landscape~\cite{Coomer2022Noise}. We cannot interpret the time at the top of the landscape as the current time, with future states of the ball known ahead of time, as cell differentiation is generally subject to randomness. In general, we are therefore unable to predict the landscape, and future states of the random system can only be known with any certainty once the process is complete, giving a historical account of the process that occurred. Moreover, we cannot obtain the two-dimensional landscape from the one-dimensional trajectories of cell differentiation subject to randomness. Thinking of the process of cell fate specification as a random dynamical system shows that it is unrealistic to interpret literally some of the landscape features. We depart from the convenient idealization of a low-dimensional landscape of cell fate, preferring to minimize assumptions in order to distil structural properties that would otherwise be missed by simpler frameworks. This perspective emphasizes the potential relevance of different properties -- for instance, of more diversely defined notions of attractors -- and makes transient dynamics and their asymptotic behaviour legible. It is also important to note that highly simplified models are likely to have only asymptotic stability, as any long transients may only arise with the more complex dynamics.

It is generally very difficult, or impossible, to find a mathematical description of the evolution function or cocycle for a specific (not necessarily random) dynamical system: this reflects the same difficulty in finding solutions for the differential equations that generate dynamical systems. The state spaces can also be difficult, or impossible, to determine, even assuming all relevant state variables are known and quantifiable. The aim of dynamical systems theory, however, is to circumvent these difficulties by providing a framework for developing and applying mathematical theory, such as the notion of an attractor, that can reveal the fundamental qualitative properties of the dynamics for real systems. Therefore, there is no need for mathematical descriptions of, for example, the state transitions. In fact, our interest in dynamical systems is usually with respect to understanding the \emph{structure} of the solutions: dynamical systems theory is primarily concerned with qualitative, rather than quantitative, properties of dynamical systems.

While dynamical systems theory is abstract, it provides a different language with which to ask more precise questions, and hopefully access a better understanding of cell fate. For instance, we may want to determine a minimal state space for representing the trajectory of a given cell from stem to terminally differentiated, which would let us identify the conditions enabling differentiation -- whether molecular, physical, random, etc. -- and locate the source of those conditions as, for instance, internal to the cell (intrinsic), or arising within the integrated cell--microenvironment system (systemic). This would let us be more precise about knowing the extent to which cell fate may -- or may not -- be predicted from any given state. In other words, this lets us talk about the uncertainties associated with the future fate of a cell given its present state, and with the necessary conditions to reach a future fate. We already know that the process of cell fate determination cannot be described in full as a Markov process~\cite{Stumpf2017Stem}. To better understand cell fate, we would like to identify the conditions that reduce uncertainties about a given fate, that is, conditions that bring the necessarily random dynamics of cellular differentiation closer to a Markov process. This is only possible within frameworks that do not assume predetermined answers to those questions; the theory of random dynamical systems, as presented here, being one of those.

\begin{quote}
    \emph{There's no sense in being precise when you don't even know what you're talking about} -- John von Neumann
\end{quote}

\end{document}